\documentclass[amsmath,amssymb,ams, preprint, superscriptaddress,nofootinbib]{revtex4-2} 
\usepackage[a4paper,textwidth=17.9cm,textheight=26cm]{geometry}
\usepackage[english]{babel}
\usepackage{ulem}
\usepackage{siunitx} 

\addto\captionsenglish{\renewcommand{\figurename}{\textbf{Fig.}}}
\addto\captionsenglish{\renewcommand{\tablename}{\textbf{Tab.}}}

\usepackage{placeins}  
\usepackage{float}

\usepackage{upgreek}
\usepackage{graphicx}

\usepackage[T1]{fontenc}
\usepackage{lmodern}
\usepackage{helvet}

\usepackage{lineno}

\usepackage{xr}
\makeatletter
\newcommand*{\addFileDependency}[1]{
  \typeout{(#1)}
  \@addtofilelist{#1}
  \IfFileExists{#1}{}{\typeout{No file #1.}}
}
\makeatother

\usepackage[font=small,labelfont=bf,format=plain,justification=justified]{caption}
\usepackage{makecell}
\usepackage{xcolor}
\usepackage{pifont}

\let\oldding\ding
\renewcommand{\ding}[2][1]{\scalebox{#1}{\oldding{#2}}}

\begin{document}
\title{Reconfigurable Single-Ring Photonic Molecule on Lithium Niobate}

\author{Tianyi Zhang}

 \affiliation{Hybrid Photonics Laboratory, École Polytechnique Fédérale de Lausanne (EPFL),  CH-1015, Switzerland}
 \affiliation{Center for Quantum Science and Engineering (EPFL), CH-1015, Switzerland}

\author{André Garcia Primo}

\affiliation{Hybrid Photonics Laboratory, École Polytechnique Fédérale de Lausanne (EPFL),  CH-1015, Switzerland}
 \affiliation{Center for Quantum Science and Engineering (EPFL), CH-1015, Switzerland}

\author{Jiawen Liu}

 \affiliation{Hybrid Photonics Laboratory, École Polytechnique Fédérale de Lausanne (EPFL),  CH-1015, Switzerland}
 \affiliation{Center for Quantum Science and Engineering (EPFL), CH-1015, Switzerland}

\author{Aleksei Gaier}

 \affiliation{Hybrid Photonics Laboratory, École Polytechnique Fédérale de Lausanne (EPFL),  CH-1015, Switzerland}
 \affiliation{Center for Quantum Science and Engineering (EPFL), CH-1015, Switzerland}

\author{Ileana-Cristina Benea-Chelmus}
\affiliation{Hybrid Photonics Laboratory, École Polytechnique Fédérale de Lausanne (EPFL),  CH-1015, Switzerland}
 \affiliation{Center for Quantum Science and Engineering (EPFL), CH-1015, Switzerland} 
\date{\today}

\begin{abstract}
Resonant photonic structures enable optical enhancement and spectral filtering and are essential for lasers, quantum emitters, transducers, or modulators. Photonic molecules, formed by mode hybridisation in two coupled resonators, break the equidistant frequency spacing of zero-dispersion resonators and provide control over their spectrum. Reconfigurability over these devices is a key asset, allowing to align photonic resonances to target frequencies on-demand. While electro-optic materials such as thin-film lithium niobate (TFLN) have enabled frequency tuning beyond traditional thermo-optic effects, they require continuous bias, posing challenges to scalability. Here we demonstrate an optically programmable, erasable, and rewritable photonic molecule realized within a single TFLN racetrack resonator. A long-lasting photorefractive grating induced through interference of co-propagating dark and bright transverse modes promotes their hybridisation, forming a single-ring photonic molecule. We observe GHz-scale hybrid-mode splitting over a 700~GHz-wide optical bandwidth and hour-long lifetimes, and show that their coupling strength can be programmed by the optical pump used to write the grating. 
By selectively pumping orthogonal hybridised modes, we further demonstrate multiple reversible all-optical write–erase–rewrite cycles of these gratings. Finally, we use this technique to realize single-sideband mmWave transduction around 107~GHz with a 5~GHz tuning bandwidth. These results establish photorefraction as a reliable mechanism for reconfigurable resonances in TFLN, and suggest a route towards tunable microwave-optical 
functionalities within a reduced footprint.
\end{abstract}
\maketitle

\section{Introduction}
As integrated photonic circuits evolve toward larger-scale and increasingly multifunctional architectures, reconfigurable and programmable control over device parameters has also become essential.
Such reconfigurability is especially important for optical resonators, whose spectral selectivity enables filtering and whose field enhancement supports nonlinear optics, modulation, and frequency transduction~\cite{shekhar_roadmapping_2024,bogaerts_programmable_2020,perez-lopez_multipurpose_2020,churaev_heterogeneously_2023,li_high_2023,Zhu:21}. However, in the absence of external tuning mechanisms, the usable bandwidth of such devices is fixed and typically restricted to a single resonance linewidth, on the order of hundreds of MHz~\cite{Xue:22, zhang2017monolithic, Siliconmicroring}. This constraint limits their applications where matching to an a priori unknown or dynamically tuned frequency is required, such as in quantum interconnects~\cite{Multani2026, mckenna_cryogenic_2020, Warner2025}, radiometry~\cite{SantamariaBotello:18}, and signal multiplexing~\cite{hsu_-chip_2023}.

Thin-film lithium niobate on insulator (TFLN) is a leading platform for resonant photonics, combining low propagation loss and strong second-order optical nonlinearities. 
Advances in microfabrication have enabled TFLN waveguides with losses below $ \SI{3}{\dB/\meter}$~\cite{nano8110910} and mm-scale resonators with intrinsic quality factors exceeding $20$ million~\cite{Zhu2024, Khalatpour2025}. 
At the device level, racetrack and microring resonators have enabled a broad range of applications, from high-speed electro-optic modulation~\cite{wang2018integrated,zhu2022spectral} and wavelength filtering to frequency-comb generation~\cite{zhang2019broadband, Zhu:21, hu_high-efficiency_2022, song_hybrid_2025} and coherent frequency transduction~\cite{Arnold2025,Multani2026,Warner2025,wang2023quantum}. 
Despite this progress, reconfigurability in TFLN still relies predominantly on electro-optic or thermo-optic tuning. Maintaining a programmed state therefore requires continuous bias or heating, and large-scale integration introduces challenges in electrode routing, power consumption, and control complexity~\cite{Guarino2007, Liu:20}. 
A reversible and long-lived mechanism for programming resonant optical responses in situ would therefore be highly desirable for scalable TFLN photonics.

Apart from designing the resonator, one of the most widely used strategies for tailoring the response of photonic devices is through hybridisation of optical resonances. This is typically achieved by evanescently coupling resonators with  matching resonance frequency, allowing light to couple electromagnetically between the two, giving rise to so-called photonic molecules~\cite{PhysRevLett.81.2582, rakovich2010photonic,Zhang2019ProgrammablePM, Peng:12, PhysRevA.96.043808}. These structures break the equidistant frequency ladder of dispersion-engineered resonators, and are essential for controlling the optical energy flow in experiments such as the laser cooling and control of acoustic modes~\cite{PhysRevA.90.053841, Thompson2008}, low-power generation of electro-optic~\cite{hu_high-efficiency_2022} and soliton frequency combs~\cite{helgason_surpassing_2023}, and single-sideband up- or down-conversion~\cite{hu_-chip_2021}. While versatile, photonic-molecule architectures typically have restricted coupling topologies, with coupling strengths largely fixed after fabrication. Therefore, most reconfigurable implementations rather tune the resonance mismatch between the coupled resonators~\cite{Warner2025}, instead of directly changing their coupling rate. The difference is that, in the former case, light is localized in either of the two cavities instead of being evenly distributed across both~\cite{AliMiri2019}. This localization gives rise to optical imbalance and degrades the system’s nonlinear response. Instead, direct control of the coupling strength is preferable because it tunes the mode splitting while preserving complete hybridisation.

An alternative to two-resonator photonic molecules is to realize photonic molecules within a single multimode resonator, where coupling occurs between different transverse modes. Recent work has established ``single-ring photonic molecules'' in multimode microrings based on bright and dark transverse modes~\cite{Lu2026MultimodeSingleRing}. Tunable or reconfigurable mode splitting has been explored in both silicon and photosensitive chalcogenide resonators~\cite{Tao2024VersatilePM,Shen2020ReconfigurableSplitting,Xia2025ReconfigurableChG}. 
The silicon implementation uses engineered bend-induced spatial-mode interaction together with active thermal tuning~\cite{Tao2024VersatilePM}, whereas the chalcogenide implementation relies on material-specific photosensitive Bragg-gratings coupling counter-propagating modes~\cite{Shen2020ReconfigurableSplitting,Xia2025ReconfigurableChG}. Lithium niobate, on the other hand, offers a distinct route through photorefraction, in which photo-excited charge carriers redistribute under internal electric fields and modify the refractive index through the linear electro-optic effect~\cite{1985ApPhA..37..191W,peterson1964electro}. 
In high-$Q$ TFLN microresonators, this process is most commonly observed as a resonance blue shift during optical pumping~\cite{Jiang:17}. 
While typically regarded as a parasitic effect, photorefraction can also imprint long-lived spatial refractive index modulations when the intracavity optical field exhibits longitudinal interference patterns. Previous studies have shown that such photorefraction-induced gratings can couple counter-propagating modes~\cite{Xu:21bragg,PhysRevLett.127.033902} and enable quasi-phase matching for nonlinear processes~\cite{Hou_2024}. These results suggest that photorefraction can be harnessed as a route to optically-controlled reconfigurability in lithium niobate resonators.

In this work, we utilize photorefraction to realize a reconfigurable single-ring photonic molecule in a multimode x-cut TFLN racetrack resonator. By optically pumping weakly coupled resonances (TE$_0$ and TM$_0$) we reinforce their hybridisation through the formation of a photorefractive grating. In contrast to previously reported photorefractive gratings, which primarily couple counter-propagating modes, we couple co-propagating modes. This satisfies phase-matching over significantly larger optical bandwidths, enabling hybridisation across multiple longitudinal resonances. We observe GHz-scale mode splitting and find that the splitting remains even after the write beam is removed, with hour-scale relaxation times. We further demonstrate that the coupling strength can be tailored through the optical write power, and the grating can be reversibly erased and rewritten by selectively exciting orthogonal hybrid modes. The written photorefractive grating thus functions as a reconfigurable intracavity coupling element. Combining the hybridised spectrum with the large electro-optic response of x-cut TFLN, we demonstrate tunable mmWave transduction around 107~GHz with a tuning bandwidth of 5~GHz, an order of magnitude larger than the optical linewidth of our resonators. These results establish photorefraction as a persistent and flexible mechanism for programming resonant optical responses in lithium-niobate photonics.

\begin{figure}[t!]
  \centering
  \includegraphics[width=16cm]{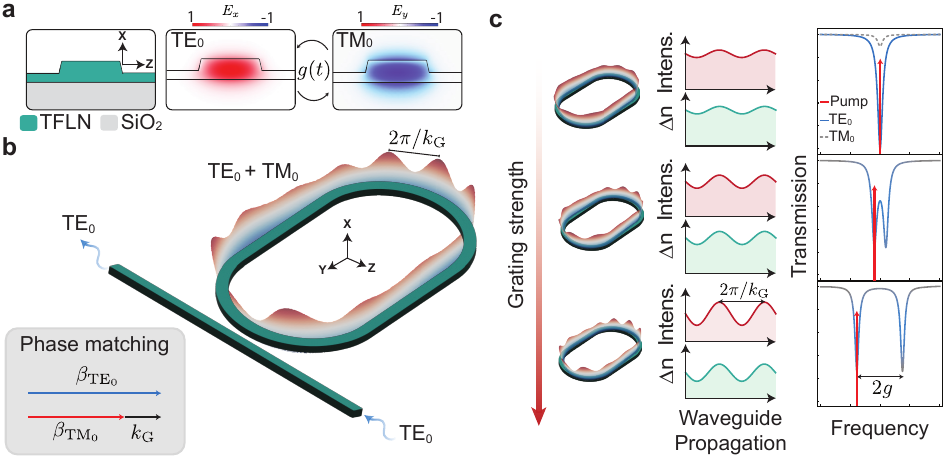}
  \renewcommand{\baselinestretch}{1} 
  \caption{\textbf{Operating principle of the single-ring photonic molecule based on an optically written photorefractive grating.}
  (a) The photonic molecule exploits the hybridisation of TE$_0$ and TM$_0$ modes with simulated mode profiles as shown.
  (b) Device concept: the bright mode (TE$_0$) is directly excited from the bus waveguide, while the dark mode (TM$_0$) is populated through intermodal coupling primarily arising along the bends. A weak initial hybridisation occurs even in the absence of optical pumping due to non-orthogonality of these modes in the bent regions. The intracavity beating of TE$_0$ and TM$_0$ modes (visualized through the intensity pattern displayed along the racetrack) writes a photorefractive grating with spatial frequency $k_\text{G} = \beta_{\text{TE}_0}-\beta_{\text{TM}_0}$.
  (c) When the participating modes are nearly degenerate (upper panel), the excitation of either the symmetric or the asymmetric hybridised mode (visualized by the red arrow) introduces a beating pattern and, consequently, a photorefractive grating which modulates the refractive index ($\Delta n$) along the racetrack (see left and central panels). Subsequent pumping of the same mode strengthens the grating (evolution top to bottom) and increases the hybrid-mode splitting $2g$.}
  \label{fig:fig1}
\end{figure}

\section{Results}
\subsection{Device concept and working mechanism}

Our single-ring photonic molecule is implemented using a racetrack resonator with a nominal waveguide width of $2~\mu\mathrm{m}$ which supports multiple mode families, including $\text{TE}_0$ and $\text{TM}_0$ (Fig.~\ref{fig:fig1}a). Optical access to the device is provided by focusing grating couplers designed for excitation of the fundamental $\text{TE}_0$ mode and a bus waveguide engineered to efficiently excite the $\text{TE}_0$ mode inside the racetrack (Fig.~\ref{fig:fig1}b). By definition, TE$_0$ then represents a bright mode of the racetrack with propagation constant $\beta_{\text{TE}_0}$. The TM$_0$ mode is not excited via the bus waveguide and consequently is a dark mode of the racetrack with propagation constant $\beta_{\text{TM}_0}$. The racetrack incorporates Euler bends where the changing propagation direction, crystal-axis projection, and bend-induced polarization mixing lead to non-orthogonality of $\text{TE}_0$ and $\text{TM}_0$, resulting in their weak hybridisation even in the absence of any optical power. A summary of the key device dimensions is provided in the Supplementary Information Note~2.

We now explain the step-by-step formation and the optical control of the photorefractive grating (Fig.~\ref{fig:fig1}b and c). Under optical pumping, both mode families, weakly hybridised, are excited and simultaneously propagate inside the cavity. Their longitudinal interference creates a spatially periodic intensity pattern along the circumference of the racetrack resonator with wavevector $k_G=\beta_{\text{TE}_0}(\omega_\text{p})-\beta_{\text{TM}_0}(\omega_\text{p})$, where $\omega_\text{p}$ is the pump frequency. The intensity pattern features stronger modulation along the bends and weaker in the straight sections of the racetrack, as shown in Fig.~\ref{fig:fig1}b. Through optical absorption, this intensity pattern excites new charge carriers. These carriers redistribute within the built-in TFLN internal fields and become trapped near the waveguide sidewalls, generating a quasi-static electric field inside the waveguide that locally modifies its refractive index through the linear electro-optic effect. This process builds up a persistent refractive-index modulation, i.e. a photorefractive grating, with the same longitudinal periodicity as the optical beating pattern $k_\text{G}$. The induced permittivity perturbation is described by
\begin{equation}
  \Delta\varepsilon(\mathbf r_\perp,s,t)
  =
  a_{\text{TE}_0}a_{\text{TM}_0}^{\ast}
  C(\mathbf r_\perp,s,t)
  e^{ik_G s}
  + \mathrm{c.c.},
  \label{eq:delta_eps}
\end{equation}
where $a_{\text{TE}_0}$ and $a_{\text{TM}_0}$ are the intracavity amplitudes, $s$ is the arclength along the racetrack, $\mathbf r_\perp$ denotes the transverse coordinates, and \(C(\mathbf r_\perp,s,t)\) is a slowly varying photorefractive grating envelope. This grating then mediates the coupling between the two mode families transforming initially weakly interacting resonances into a hybridised photonic molecule with coupling strength $g$. When the modes in question are nearly degenerate, the observable hybrid-mode splitting is approximately $2g$, directly reflecting the intermodal coupling strength, as illustrated in Fig.~\ref{fig:fig1}c. With repeated pumping, the strength of the written photorefractive grating increases (see grating evolution on the left and intensity/refractive index modulation in central panel of Fig.~\ref{fig:fig1}c, top to bottom), leading to an optically controlled $g$ and hence hybrid-mode splitting $2g$ (right panel Fig.~\ref{fig:fig1}c, top to bottom).

This system can be modeled as two coupled optical modes with a total intermodal coupling rate given by $g(t)=g_0+g_{\mathrm{PR}}(t)$, where $g_0$ is the weak static coupling present before optical writing and $g_{\mathrm{PR}}(t)$ is the optically induced photorefractive contribution. The coupling approaches a steady-state value on the timescale of the photorefractive response. To account for this effect, we model the coupling rate as $g_{\text{PR}}(\omega_\text{p},t)\propto 
\left(1-e^{-t/\tau_{\text{PR}}}\right)
a_{\text{TE}_0}(\omega_\text{p})
a_{\text{TM}_0}^{\ast}(\omega_\text{p})$, where $\tau_{\mathrm{PR}}$ is the characteristic photorefractive response time and $a_{\text{TE}_0}(\omega_\text{p})$, $a_{\text{TM}_0}(\omega_\text{p})$ are the intracavity bright- and dark-mode amplitudes at the pump frequency. A full two-mode temporal coupled-mode model, the corresponding transmission function, and the derivation of the grating-induced coupling and bandwidth expressions are given in Supplementary Information Note~1.  

\subsection{Properties of the optically written photonic molecule}

\begin{figure}[t!]
  \centering
  \includegraphics[width=16cm]{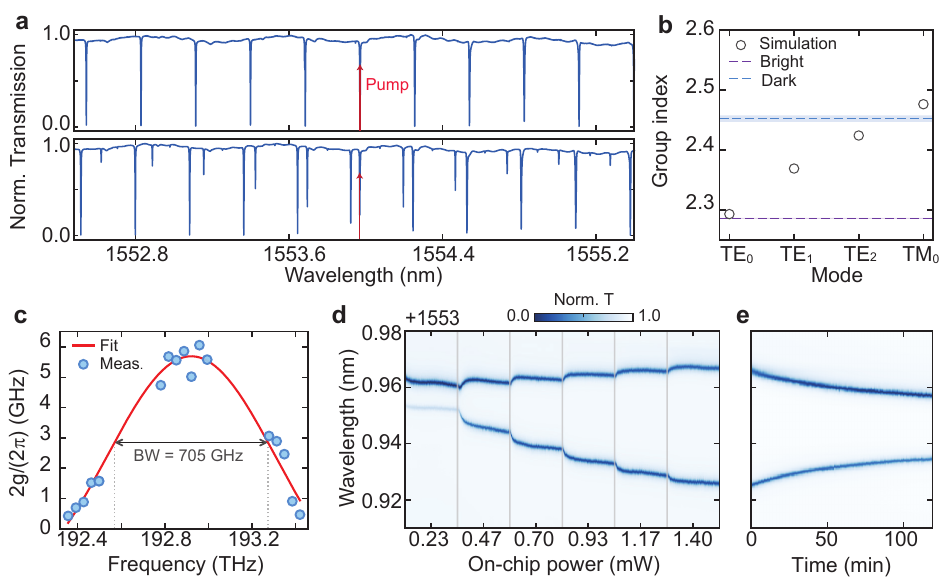}
  \renewcommand{\baselinestretch}{1} 
  \caption{\textbf{Properties of photorefractive grating: optical bandwidth, power dependence and relaxation times.}
  (a) Normalized transmission spectra measured before pumping (upper panel) and after cumulative pumping (lower panel), showcase the emergence of hybrid doublets for more than 11 longitudinal modes around the pumped resonance. 
  (b) Comparison between the group indices of the bare modes extracted from the resonance positions and the simulated values of all higher order modes supported by the structure. We assign the TE$_0$ and TM$_0$ modes as the bright and dark modes participating in formation of the single-ring photonic molecule proposed here. The TE$_2$ mode is excluded through a Q-factor argument, see text. 
  (c) The optical bandwidth of the photorefractive grating is extracted by analyzing the coupling strength of the various doublets versus wavelength. Fitting the analytically predicted sinc function to the experimental data allows extracting a full-width half maximum of 705~GHz.
  (d) Pump-power dependence of hybrid-mode splitting. The evolution of the transmission spectrum as the on-chip write power is increased from 0.233~mW to 1.398~mW in steps of 0.233~mW and reveals saturation effects that happen increasingly faster for larger pump powers. The device is pumped for 100 iterations at each power setting.
  (e) A relaxation of the hybrid-mode splitting is observed after turning off optical pumping by tracking the transmission of the photonic molecule as a function of time. The relaxation time constant is approximately 65 minutes.}
  \label{fig:fig2}
\end{figure}

We now demonstrate the operation of our device as a single-ring photonic molecule. A low-power, fast-scan transmission spectrum reveals the resonant frequencies of the TE$_0$ mode prior to optical pumping (upper panel of Fig.~\ref{fig:fig2}a). Among the available longitudinal resonances, we select a TE$_0$--TM$_0$ pair that is close to degeneracy. Tuning the chip temperature allows to fulfill this condition for any of the mode pairs, owing to the different thermo-optic tuning rates of the two mode families (see Supplementary Information Note~6). We pump the corresponding hybrid resonance, marked by the red arrow in the upper panel of Fig.~\ref{fig:fig2}a, thereby realizing the near-degenerate writing condition illustrated in the upper panel of Fig.~\ref{fig:fig1}c, where $\omega_\mathrm{p}$ initially coincides with $\omega_{\mathrm{TE}_0}\sim\omega_{\mathrm{TM}_0}$. After cumulative pumping (the protocol is described in the Methods), hybrid-mode doublets emerge around the pumped resonance due to photorefraction-induced intermodal coupling, as observed in the transmission spectrum after pumping shown in the lower panel of Fig.~\ref{fig:fig2}a.

To confirm the hybridisation of TE$_0$ and TM$_0$ and rule out any other possibilities (e.g. TE$_0$ and TE$_1$ or TE$_0$ and TE$_2$), we extract the group indices of the measured mode families ($n_{g,\mathrm{TE_0}} = 2.286 \pm 0.001$ and $n_{\text{g},\mathrm{TM_0}} = 2.452 \pm 0.005$, where the uncertainties denote the standard deviation over the measured resonances) over multiple free spectral ranges using the low-power spectrum acquired before grating formation. These values are compared in Fig.~\ref{fig:fig2}b with numerical eigenmode simulations, in which the angle-dependent group index on the x-cut TFLN platform is averaged over the racetrack optical path to account for the varying crystal-axis projection along the resonator. Details about the procedure used to extract the experimental values can be found in Supplementary Information Note~3. We attribute the measured group index of the bright mode to TE$_0$ and of the dark mode to TM$_0$. The ambiguity between the TM$_0$ and TE$_2$ modes in the dark mode identification is resolved because of their vastly different quality factors. Specifically, coplanar gold transmission lines are patterned next to the straight section of our racetrack to enable millimeterwave-optical transduction (discussed later in the manuscript). The metal introduces significant absorption on the TE$_2$ mode, which is less confined to the waveguide core, resulting in  absorption-limited quality factors far inferior to those experimentally observed for the dark mode (see Supplementary Information Note~3).

\FloatBarrier

Having established the nature of coupling, we proceed to quantify the optical bandwidth of a single inscribed grating. The coupling bandwidth is important when e.g. envisioning operating a photonic molecule across several longitudinal resonances. Unlike coupling of counter-propagating modes reported in literature, the co-propagating nature of our approach allows for comparably relaxed phase matching conditions, supporting a much broader coupling bandwidth that extends over more than 23 longitudinal resonances. The phase matching amplitude decays away from the pump frequency used to write the grating following a sinc-shaped envelope with a bandwidth $\Delta\omega \approx \frac{7.58\,c}{|\Delta n_\text{g}|L_{\mathrm{eff}}}$ (derivation in Supplementary Information Note~1), where $\Delta n_\text{g}=n_{\text{g},\mathrm{TE_0}}-n_{\text{g},\mathrm{TM_0}}=0.166 \pm 0.005$ is the group-index difference between the two mode families and $L_{\mathrm{eff}}$ is the effective interaction length of the written grating. Engineering $\Delta n_g$ allows extending the bandwidth further. 

To quantify the optical bandwidth of the written coupling, we fit each measured doublet around the
pump frequency with a coupled-mode model and extract \(g\) for each resonance pair (fitting details in Supplementary Information Note~4). The resulting
\(2g\) values are shown in Fig.~\ref{fig:fig2}c as a function of the frequency of the mode pair
and are well described by a sinc envelope
$2g(f)=\,2g_0\,\mathrm{sinc}\!\left[\frac{\Delta n_\text{g} L_{\mathrm{eff}}}{c}\,\pi(f-f_\text{p})\right]$ (see Supplementary Information Note~1),
with fixed \(\lambda_\text{p} = 1553.965~\mathrm{nm}\) determined experimentally.
Using the independently extracted group-index difference \(\Delta n_\text{g} \), the fit yields \(2g_0/2\pi=(5.69\pm0.15)~\mathrm{GHz}\) and a full-width half maximum coupling bandwidth
\(\Delta f_{\mathrm{bw}}^{(\mathrm{fit})}=0.71\pm0.02~\mathrm{THz}\). From the bandwidth expression introduced
above, we infer an effective interaction length \(L_{\mathrm{eff}} \approx \frac{7.58\,c}{2\pi |\Delta n_\text{g}| \Delta f_{\mathrm{bw}}^{(\mathrm{fit})}} \approx 3.09~\mathrm{mm} \approx 0.84\,L\), shorter than the geometric
racetrack perimeter. This is consistent with a non-uniformly written photorefractive grating whose longitudinal weighting is
stronger in sections where TE\(_0\)-TM\(_0\) beating and photorefractive writing are stronger (in the bends).

We now perform a systematic study of the dependence of hybrid-mode splitting on pump power. We set the pump power in the bus waveguide to range 0.233~mW to 1.398~mW in steps of 0.233~mW. For each power setting, we pump the high-wavelength resonance for 100~iterations using the pump-measure protocol described in the Methods~(Fig.~\ref{fig:fig2}d). We observe a saturated splitting for each pump power, providing a knob for optically programmable control of hybrid-mode splitting in single-ring photonic molecules. At high pump powers, the increase in saturated hybrid-mode splitting becomes sublinear, revealing bounds on the maximum achievable splitting in a given device. A more in-depth analysis is given in the Supplementary Information Note~5.

Lastly, for practical applications, it is important that the inscribed grating is persistent over laboratory timescales. We now proceed to demonstrate hour-long decay times of our grating after pumping is switched off (Fig.~\ref{fig:fig2}e). We capture the leading behavior over the first two hours by tracking the transmission spectra of the photonic molecule and fitting the extracted $2g$ with a model containing a dominant decay and a residual offset $2g(t)= 2g_\infty + A e^{-t/\tau},$ where $\tau$ describes the dominant hour-scale relaxation, $g_\infty$ accounts for long-lived residual coupling, and $A$ denotes the decay amplitude of the angular-frequency splitting. The fit reveals hours-long relaxation times, with $A/(2\pi) = (2.58 \pm 0.01)$~GHz, $\tau = (65.2 \pm 0.8)$~min, and $2g_\infty/(2\pi) = (2.38 \pm 0.01)$~GHz, where the uncertainties denote one-standard-error estimates.

\subsection{Reconfigurable photonic molecule}

\begin{figure}[!tbp]
  \centering
  \includegraphics[width=16cm]{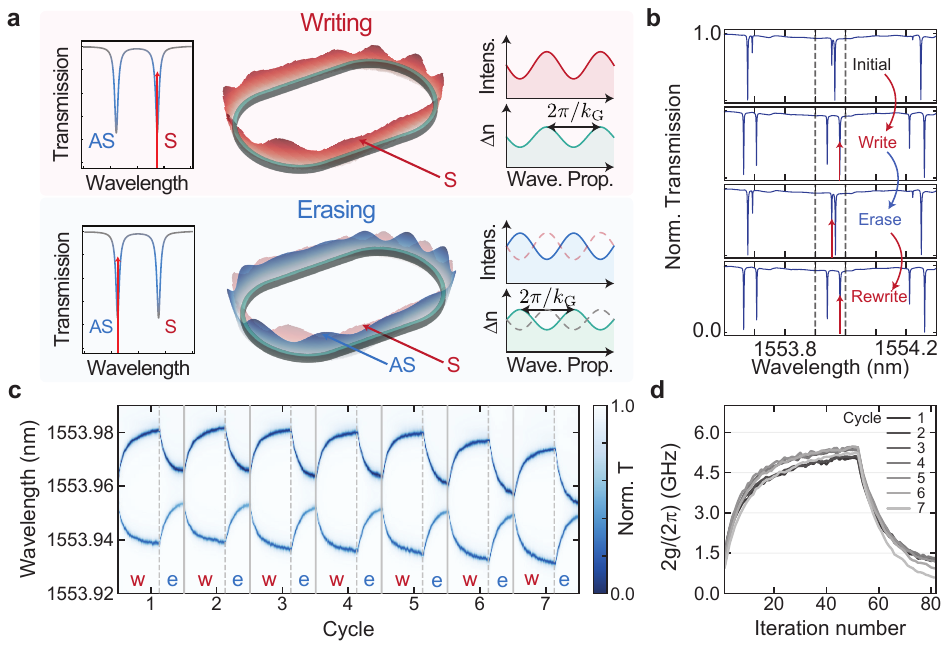}
  \renewcommand{\baselinestretch}{1} 
  \caption{\textbf{Reconfigurable write--erase--rewrite operation of the single-ring photonic molecule over multiple cycles.}
  (a) Working principle of the write--erase mechanism. We exploit the orthogonality of the symmetric (S) and antisymmetric (AS) hybrid modes which feature intra-cavity fields proportional to $a_{\text{TE}_0}+a_{\text{TM}_0}$ and $a_{\text{TE}_0}-a_{\text{TM}_0}$, respectively. The $\pi$-shift between the bright (TE$_0$) and dark components (TM$_0$) introduces a spatial shift of half a grating period when comparing the intensity distribution of the two orthogonal hybrid modes (see alignment between maxima and minima in the schematics and the refractive index distributions). Pumping the two hybrid modes in succession enables an initial write step, followed by the cancellation of the previously written grating.
  (b) Representative transmission spectra showing the initial state, the written state, the erased state, and the rewritten state. The red arrows indicate toggling between the symmetric and antisymmetric modes as a mean to swap from writing to erasing and vice-versa. 
  (c) Repeated write--erase--rewrite cycles over time show reproducible behavior over several cycles, where ``w'' and ``e'' denote the writing and erasing steps, respectively. (d) The extracted coupling strength $2g$ is overlaid for the 7 cycles, allowing to establish a 1:1 correspondence of the iteration number and the achieved coupling strength.}
  \label{fig:fig4}
\end{figure}

In the following, we show that the written hybrid-mode splitting can be reversibly erased and reprogrammed over multiple cycles. This reconfigurability is essential for applications requiring precise frequency alignment e.g. for resonant transduction. To achieve this, we exploit the orthogonality of the two hybridised modes which are symmetric and antisymmetric superpositions of the TE$_0$ and TM$_0$ modes with corresponding intracavity fields $a_{\text{TE}_0}+a_{\text{TM}_0}$ and $ a_{\text{TE}_0}- a_{\text{TM}_0}$~\cite{AliMiri2019}. Because these two supermodes differ by a phase of $\pi$ between their bright and dark components, their longitudinal intensity beat patterns and the refractive index modulation $\Delta n$ are shifted by half a grating period along the waveguide (corresponding to replacing $ a_{\text{TE}_0}a_{\text{TM}_0}^{\ast}\rightarrow-a_{\text{TE}_0}a_{\text{TM}_0}^{\ast}$ and consequently $e^{ik_G s}
  \rightarrow
  -e^{ik_G s}
  =
  e^{ik_Gs+\pi}$ in Eq.~\ref{eq:delta_eps}), as illustrated in Fig.~\ref{fig:fig4}a. Pumping the two supermodes in sequence therefore writes index modulations with opposite sign and equal spatial frequency $k_\text{G}$ on top of each other, leading to a partial cancellation and eventual erasure of the initially written grating. This enables all-optical write--erase--rewrite operation of the intermodal coupling over many cycles.

Optical spectra resulting from write--erase--rewrite cycles of the photorefractive gratings are shown in Fig.~\ref{fig:fig4}b (red arrow indicates the pumped resonance). To ensure consistent pumping on either the symmetric or antisymmetric mode, we chose to pump a pair where the detuning between TE$_0$ and TM$_0$ is near-zero but not equally zero, allowing their unambiguous tracking via the resonance frequency. An initial photorefractive grating is inscribed by first pumping the devices on the higher-wavelength hybrid mode for 50 iterations, using a pump-measure protocol described in the Methods. We observe a progressive enhancement of the hybrid-mode splitting, followed by the saturation we discuss in Fig.~\ref{fig:fig4}d. We subsequently pump the lower-wavelength hybrid mode for 30 iterations to erase the previously formed grating. We observe a rapid reduction of the hybrid-mode splitting, followed once again by saturation dynamics.  We repeat this procedure for seven consecutive cycles, corresponding to a total measurement time of approximately $10$~h. By plotting the coupling strength $2g$ as a function of iteration for the 7 cycles we find that a 1:1 correspondence between iteration number and coupling strength can be established (Fig.~\ref{fig:fig4}d).

These results show that the photorefractive grating can be written, erased, rewritten, and programmed with good repeatability over many cycles. The hybrid-mode splitting faithfully follows the programmed state throughout the sequence. Despite the small drift of the resonance wavelength due to sample temperature fluctuations and photorefractive build-up, the reversibility of the splitting is preserved, indicating that photorefraction-induced intermodal coupling in this platform provides a robust and reconfigurable degree of freedom.

\FloatBarrier

\subsection{Tunable mmWave transduction based on reconfigurable photonic molecule}

\begin{figure}[!t]
  \centering
  \includegraphics[width=16cm]{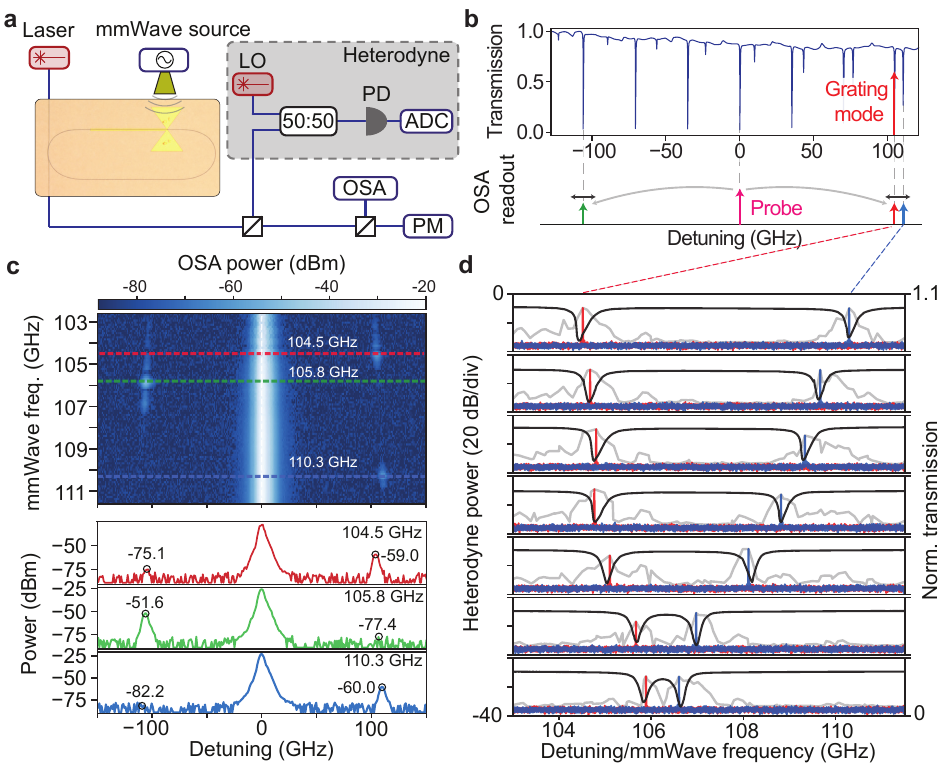}
  \renewcommand{\baselinestretch}{1} 
  \caption{\textbf{Tunable single-sideband transduction of  mmWaves based on single-ring photonic molecule.}
  (a) Experimental configuration. Wireless mmWaves are coupled to the chip via an antenna-coupled transmission line. The mmWaves are up-converted to the telecom band, creating sidebands, exploiting the $r_{33}$ electro-optic coefficient of TFLN. The transmitted light is analyzed with a power meter (PM), an optical spectrum analyzer (OSA), or by heterodyne detection mixing with a swept local oscillator (LO) on a photodiode (PD) followed by an analog-to-digital converter (ADC).
  (b) Transmission spectrum after inscribing a photorefractive grating through pumping the resonance marked with the red arrow. Laser light at a wavelength corresponding to the pink arrow (three FSRs away) is coupled to the racetrack to undergo frequency mixing with the mmWaves. Green, red and blue arrows mark hybrid modes targeted by mmWave sideband generation, underlining their uneven spacing.
  (c) Top: OSA spectrogram versus mmWave frequency confirms efficient single-sideband transduction at three mmWave frequencies (dashed lines at 104.5, 105.8 and 110.3 GHz, corresponding to the one blue-detuned and the two red-detuned frequencies shown in (b)). Bottom: linecuts at these three frequencies, color-matched to the target modes in (b) showcase a sideband suppression of at least 15~dB.
  (d) Tunable mmWave transduction is demonstrated by inscribing gratings with successively increasing strength (bottom to top), as visible by the increasingly large splitting in the transmission spectra of the hybrid doublet (black). For each $2g$, an up-conversion experiment is performed where the mmWave frequency is swept across the entire range from 103-111 GHz. Plotting the corresponding sideband upconversion envelope (grey, obtained by tracking the sideband peak in the heterodyne spectrum) reveals successful up-conversion when the incoming mmWave is tuned to the lower- and higher-frequency hybrid mode, respectively. Red and blue traces are the maximum-power sidebands.}
  \label{fig:fig5}
\end{figure}

We next use our proposed single-ring photonic molecule to demonstrate tunable mmWave transduction using the scheme shown in Fig.~\ref{fig:fig5}a. Wireless mmWaves are coupled to the chip via an on-chip antenna-coupled transmission line, patterned along the straight section of the racetrack. This enables frequency up-conversion of the mmWaves to the telecom via the $r_{33}$ electro-optic coefficient of TFLN, similar to our recent work~\cite{gaier2025}. The transmission line design follows the general geometry reported in Ref.~\cite{Lampert2025}.  The device here is identical to the one used in all experiments described in the sections above except for the added on-chip bowtie antenna at the end of the transmission line for wireless mmWave coupling. We use a tunable mmWave source emitting wireless radiation in the W band. The transmitted optical signal is analyzed in three complementary ways: with a power meter for resonance tracking, an optical spectrum analyzer to resolve the sideband structure, and by heterodyne detection for high-resolution spectral analysis (see Methods for details).

In our transduction measurement, we first write the photorefractive grating by pumping a selected hybrid resonance at a chosen write power until the splitting saturates. We refer to this mode as the grating mode and indicate it by the red arrow in Fig.~\ref{fig:fig5}b. The same laser is then tuned to a resonance three FSRs away ($\sim$106\,GHz, pink arrow in Fig.~\ref{fig:fig5}b) and used as a pump for the transduction measurement. In this setting, the optical spectrum around the pumped resonance is generally asymmetric because the detuning between bright and dark resonances varies from one longitudinal mode pair to another. Importantly, the laser power is reduced at this step to minimize its effects on the already written photorefractive grating. The readout is completed within about 3\,min, well below the dominant photorefractive relaxation time extracted in Fig.~\ref{fig:fig2}e.

The electro-optic interaction generates sidebands around the pump tone. The asymmetric spectrum of the photonic molecule is reflected in the susceptibilities of the up-converted and down-converted sidebands, opening up the possibility for single-sideband transduction. Fig.~\ref{fig:fig5}c (top) shows optical spectra of the signal transmitted through our device as a function of the mmWave drive frequency. When the mmWave frequency matches the detuning between the pump and a neighboring optical mode (on either the red or blue side), we observe a strong sideband signal. In our measurements, this happens at three different frequencies (104.5, 105.8 and 110.3~GHz), corresponding to the three highlighted optical modes in Fig.~\ref{fig:fig5}b (red, blue and green arrows). The linecuts shown in the bottom panels of Fig.~\ref{fig:fig5}c, color-matched to their respective target modes, reveal single-sideband modulation with sideband suppression of 16.1, 25.8 and 22.2\,dB, respectively.

We then use the power dependence of the photorefractive grating to tune the resonant transduction frequency. As established above, different write powers lead to different steady-state splittings. Accordingly, we write the same grating mode at different optical powers and, once the splitting has saturated, repeat the transduction measurement for each case. To resolve the upconversion sidebands more precisely, we use the heterodyne readout shown in Fig.~\ref{fig:fig5}a. For each mmWave drive frequency, we sweep a local oscillator and record its beat note with the chip output using a photodiode, yielding a high-resolution spectrum of the generated sidebands, as shown in Fig.~\ref{fig:fig5}d.  Connecting the peaks of these sidebands across the swept mmWave range defines the upconversion envelope (shown in gray), while the highest-power sidebands are highlighted in red and blue. These curves are overlaid with the doublet transmission spectrum for each write power, shown in dB and linear scale, respectively.

The sideband envelope exhibits two maxima that generally coincide with the two dips of the transmission spectrum, confirming that resonant upconversion is enhanced when the sideband aligns with a hybrid mode. As the write power is increased, both maxima shift systematically in accordance with the corresponding hybrid-mode dips, demonstrating that optical programming of the intracavity coupling directly translates into programmable tuning of the mmWave transduction frequency. A slight mismatch between the transmission curves and the upconverted spectrum may originate from non-ideal alignment of the pump frequency with the resonances of the photonic molecule, or from a slight decay of the grating strength between the measurements, requiring future investigations. Overall, the programmed hybrid spectrum defines a tunable mmWave transduction window centered near 107\,GHz with a 5~GHz total tuning bandwidth, where the upper-frequency maximum shifts by about 3.6\,GHz, while the lower-frequency maximum shifts by about 1.4\,GHz.

\section{Discussion}
In summary, we show that photorefraction in thin-film lithium niobate can be harnessed as a programmable intermodal coupling mechanism. In our device, the photorefractive response writes a long-lived and large bandwidth grating that couples two co-propagating transverse modes inside a single racetrack resonator. Such grating can be written, erased, and rewritten entirely optically, thereby realizing an optically-controlled single-ring photonic molecule. Although of different physical origin, it's worth noting that the photorefractive grating proposed here shares similarities with optically induced gratings in $\mathrm{Si_3N_4}$ resonators, such as automatically fulfilling quasi-phase matching \cite{li_down-converted_2025, zhou_self-organized_2025,nitiss_optically_2022}. Since the photorefraction-induced hybrid-mode splitting can be continuously set by its optical write power in an x-cut TFLN chip, we functionalized it for tunable mmWave and microwave transduction, surpassing the linewidth bottleneck of racetrack resonances by a factor of 20. 

We find that the write and erase dynamics are slow and therefore better suited to applications requiring long-lived reconfigurability rather than high-speed real-time tuning. The observed hour-scale relaxation is already sufficient for proof-of-principle transduction protocols, but longer-lived and more reproducible photorefractive states will be important for practical implementations. In particular, cryogenic operation may offer substantially longer photorefractive lifetimes. Furthermore, the lifetime, write efficiency, and residual coupling of the photorefractive grating are likely dependent on fabrication conditions, material quality, and defect or trap distributions, pointing towards future studies that could systematically engineer or study these aspects.

More broadly, beyond tunable mmWave and microwave transduction, such optically programmed mode splitting may be useful for optical trimming of resonance positions, local engineering of effective FSR and mode dispersion, and reconfigurable spectral shaping, with potential applications in optical filtering, nonlinear photonic circuits, programmable quantum photonics~\cite{aharonovich_programmable_2026}, and Kerr-comb-related spectral engineering. The underlying idea of programmable intracavity coupling may be extended beyond ring resonators to other cavity geometries, including photonic-crystal cavities~\cite{li2026polarization}.

\section{Methods}

\subsection{Device fabrication}\label{Meth:fab}

The devices were fabricated on an X-cut thin-film lithium niobate on insulator (LNOI) wafer, consisting of a $600$-nm-thick thin-film lithium niobate layer on a $4.7~\mu\mathrm{m}$ $\mathrm{SiO_2}$ buried oxide layer supported by a high-resistivity silicon substrate. Photonic structures were defined by electron-beam lithography using a FOX16 hard mask and developed in a 25\% TMAH solution. The pattern was transferred into the lithium niobate layer by ion-beam etching operated at $150$~W for $15$~min with a $10^\circ$ incidence angle, resulting in rib waveguides with an etch depth of approximately $350$~nm and sidewall angles of about $75^\circ$. Lithium-niobate redeposition was removed in a heated solution of 40\% KOH and 30\% $\mathrm{H_2O_2}$ (3:1 by volume) at $85^\circ$C. The coplanar transmission line and on-chip antenna were fabricated by a lift-off process using an MMA/PMMA bilayer resist. After electron-beam exposure and development, a $5$-nm Ti adhesion layer and a $300$-nm Au layer were sequentially deposited by electron-beam evaporation, followed by lift-off in acetone, rinsing in IPA, and drying under nitrogen flow.

\subsection{Experiment details and measurement protocols}\label{Meth:setup}

\subsubsection{Optical setup and power calibration}
Continuous-wave optical measurements were performed with a tunable laser source (Keysight N7776C). Light from the laser was launched into a polarization-maintaining (PM) fiber and coupled to the chip through a PM fiber array aligned to the focusing grating couplers. The fiber array was mounted on a multi-axis nanopositioning system (SmarAct) for chip alignment. All measurements were performed with TE-polarized input and output light. The transmitted optical signal from the through port was recorded with a power meter (Keysight N7742C). In the mmWave transduction measurements, an erbium-doped fiber amplifier (Nuphoton stretched MSA PM 1~W EDFA, model series EDFA-C0-PM-SMR-30-20-FCA) was used together with a fixed $10$-dB attenuator to reach the desired optical power level before injection into the chip. 

All powers reported in this work are given as estimated on-chip powers (along the bus waveguide, prior to the racetrack coupling region) and account for a pre-calibrated total chip insertion loss of $\sim 18.4$~dB, corresponding to a single-coupler loss of $\sim 9.2$~dB. Transmission spectra were acquired by sweeping the tunable laser across the resonance window of interest at $10~\mathrm{nm/s}$.

\subsubsection{Photorefractive writing, erasing and pump--probe protocol}\label{protocol}

To characterize the formation of the photorefractive (PR) grating, a transmission spectrum was first recorded before pumping. The device was then resonantly pumped at a selected cavity resonance, and the spectrum was measured again after pumping to monitor the emergence of hybrid-mode splitting. For the write-erase-write measurements in Fig.~\ref{fig:fig4} and the power-dependent measurements in Supplementary Information Note 5, an iterative pump--probe protocol was employed in order to compensate for resonance drift during the writing process. Each iteration consisted of three consecutive steps: (i) a wavelength sweep to determine the instantaneous resonance position, (ii) resonant pumping at the selected wavelength for a fixed dwell time, and (iii) a second wavelength sweep to record the updated transmission spectrum and re-center the pump on the shifted resonance. In the measurements reported in the main text, the pumping dwell time was approximately $20$~s, while the probe sweep required approximately $30$~s. The iteration index therefore provides a practical measure of the cumulative intracavity exposure time during PR writing.

\subsubsection{mmWave transduction setup}

The mmWave radiation was generated using an Anritsu microwave signal generator (MG36241A) followed by a  Virginia Diodes signal generator extension module operated in the ×9 multiplication configuration. The continuous-wave mmWave signal was radiated into free space by a horn antenna with 21 dBi gain and directed to the chip using a Polymethylpentene (TPX) lens system. Specifically, the beam was first focused by a 10 mm focal length, 1-inch diameter lens, subsequently re-collimated with a 65 mm focal length, 2-inch diameter lens, and finally focused onto the on-chip antenna with a 100 mm focal length lens. The antenna captured the incoming mmWave signal and coupled it into the coplanar transmission line. In this configuration, the guided optical mode in the resonator and the electrically driven mmWave field co-propagate in the interaction region, leading to electro-optically generated optical sidebands. The chip was mounted in a custom chip holder placed on a temperature-controlled mini-series breadboard (Thorlabs).

The optical output was characterized using an optical spectrum analyzer (Anritsu MS9740B) and, for higher spectral precision, by heterodyne readout. In the latter case, a Keysight N7742C tunable laser served as a swept local oscillator, operated in continuous-sweep mode at 2.5 THz/s. The resulting beat note was detected on a photodiode (PD100AC Koheron), recorded in the time domain by the M4i.2442-x8 digitizer from Spectrum Instrumentation, and converted to the frequency domain using the calibrated laser sweep. 
\vspace{1cm}

\textbf{Data Availability}
The measurement data generated in this study will be deposited in the Zenodo database under https://doi.org/10.5281/zenodo.19919468 prior to publication.

\medskip
\textbf{Code Availability}
The code used to plot the data within this paper will be deposited in the Zenodo database under https://doi.org/10.5281/zenodo.19919468 prior to publication.

\medskip
\textbf{Acknowledgments}
T.Z. acknowledges support from the Swiss National Science Foundation through grant number 515075 (SNSF-NSF lead agency project). A.G.P. acknowledges funding through the Excellence Postdoctoral Fellowship Programme from the Quantum Science Center at EPFL and the Swiss National Science Foundation grant number 515449 (SPARK). J.L. acknowledges funding through the Excellence Postdoctoral Fellowship Programme from the Quantum Science Center at EPFL and the Swiss National Science Foundation grant number 503051 (Swiss Quantum Call). A.G. and I.C.B.C. acknowledge funding from the European Union’s Horizon Europe research and innovation program under project MIRAQLS with grant agreement No 101070700. The chips were fabricated at the Center of MicroNanoTechnology (Cmi) at EPFL.

\medskip
\textbf{Author contributions} T.Z., A.G.P. and I.C.B.C. developed the concept of reconfigurable single-ring photonic molecule. T.Z. carried out the measurements, developed the theory and analyzed the data. A.G.P. carried out the COMSOL simulations and assisted with the theory and data analysis. J.L. designed and fabricated the samples, and assisted with the measurements and data analysis. A.G. developed the heterodyne setup for the transduction experiment. All authors discussed the results. T.Z., A.G.P. and I.C.B.C. wrote the manuscript with feedback from all authors. The work was done under the supervision of I.C.B.C.

\medskip
\textbf{Competing interests}
The authors declare no competing interests.

\textbf{Corresponding authors} Correspondence to Tianyi Zhang (tianyi.zhang@epfl.ch) or Ileana-Cristina Benea-Chelmus (cristina.benea@epfl.ch). 

\clearpage

\setcounter{figure}{0}
\setcounter{table}{0}
\setcounter{equation}{0}
\setcounter{subsection}{0}
\renewcommand{\thefigure}{S\arabic{figure}}
\renewcommand{\thetable}{S\arabic{table}}
\renewcommand{\theequation}{S\arabic{equation}}

\renewcommand{\figurename}{Supplementary Fig.}
\renewcommand{\tablename}{Supplementary Table}

\begin{center}
{\Large\bfseries Supplementary Information for}\\[0.8em]
{\large Reconfigurable Single-Ring Photonic Molecule on Lithium Niobate}\\[1em]
{\small Tianyi Zhang, André Garcia Primo, Jiawen Liu, Aleksei Gaier, I.-C. Benea-Chelmus}
\end{center}

\clearpage
\section*{Supplementary Note 1. Two-mode coupled-mode model and photorefractive coupling}
\label{sec:cmt_model}
This note derives the two-mode coupled-mode model used to fit the measured spectra, together with photorefractive grating build-up and its phase-matching bandwidth.
\subsection{Coupled-mode equations and transmission}
We define $a_b$ and $a_d$ the intracavity amplitudes of two bright and dark mode families, and $s_{\mathrm{in}}$, $s_{\mathrm{out}}$ the bus-waveguide input and output amplitudes. We use the normalization in which $|s|^2$ is the bus-waveguide power and $|a_j|^2$ ($j\in\{b,d\}$) is the energy stored in mode family $j$. 

By definition, bright describes a racetrack mode that can be accessed via the bus waveguide. Conversely, a dark mode can not be accessed via the bus waveguide. In the case presented in the main text, the bright mode corresponds to TE$_0$ while the dark mode corresponds to TM$_0$, but, more generally, any of the modes of the racetrack can become bright, or dark if engineered through the design of the coupling region from the bus waveguide to the racetrack.

The two-mode coupled-mode equations read
\begin{align}
  &\dot{a}_b = \left(i\Delta_b-\tfrac{\kappa_b}{2}\right)a_b + i g(t)\,a_d + \sqrt{\kappa_{b\mathrm e}}\,s_{\mathrm{in}}, \label{eq:tcmt_ab}\\
  &\dot{a}_d = \left(i\Delta_d-\tfrac{\kappa_d}{2}\right)a_d + i g(t)^*\,a_b, \label{eq:tcmt_ad}\\
  &s_{\mathrm{out}} = s_{\mathrm{in}}-\sqrt{\kappa_{b\mathrm e}}\,a_b, \label{eq:tcmt_io}
\end{align}
with laser-cavity detunings $\Delta_j=\omega-\omega_j$ from the resonances $\omega_j$, and total decay rates $\kappa_j=\kappa_{j\mathrm i}+\kappa_{j\mathrm e}$ summing up intrinsic (scattering and absorption) and extrinsic (coupling to the bus waveguide) contributions. The coupling amplitude \(g(t)\) contains both the residual static coupling present before optical writing $g_0$ and the contribution from the written photorefractive grating $g_{\mathrm{PR}}(t)$. 
 The total intermodal coupling is
\begin{equation}
    g(t) = g_0 + g_{\mathrm{PR}}(t),
    \label{eq:g_decomposition}
\end{equation}
 where $g_0$ is the residual static coupling present before optical writing and $g_{\mathrm{PR}}(t)$ is the photorefractive contribution. The photorefractive dynamics are much slower than both the cavity photon lifetime and the duration of an individual low-power wavelength sweep, so $g(t)$ is taken as constant during a single wavelength sweep.

In steady state, Eqs.~\eqref{eq:tcmt_ab}--\eqref{eq:tcmt_io} give the normalized transmission
\begin{equation}
  T(\omega)=
  \left|
  1-\kappa_{b\mathrm e}
  \frac{\tfrac{\kappa_d}{2}-i\Delta_d}
  {\left(\tfrac{\kappa_b}{2}-i\Delta_b\right)\left(\tfrac{\kappa_d}{2}-i\Delta_d\right)+g^2}
  \right|^2,
  \label{eq:transmission}
\end{equation}
which is the function used to fit the measured spectra (Supplementary Note 4) with $g = |g(t)|$. The hybrid eigenfrequencies follow from the poles of Eq.~\eqref{eq:transmission}. Neglecting loss-induced frequency pulling,
\begin{equation}
  \omega_\pm = \tfrac{1}{2}(\omega_b+\omega_d)
  \pm
  \sqrt{g^2+(\Delta/2)^2},
  \qquad
  \Delta=\omega_d-\omega_b.
  \label{eq:hybrid_freq}
\end{equation}

\begin{figure}[!htbp]
    \centering
    \includegraphics[width=\linewidth]{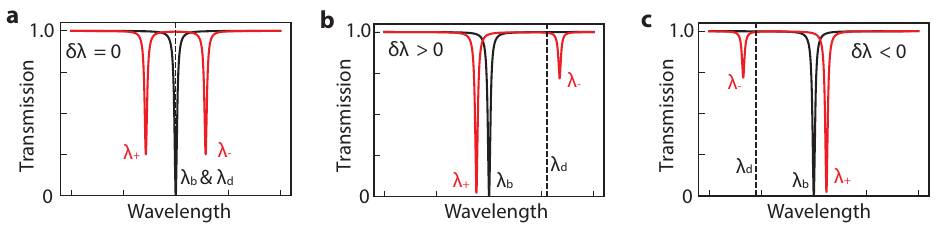}
    \caption{\textbf{Two-mode transmission for different bright--dark detunings.} Black: bare bright-mode transmission ($g=0$). Red: hybrid spectrum at finite $g$. $\lambda_b$, $\lambda_d$: bare bright/dark resonances; $\lambda_\pm$: hybrid resonances. (a) $\delta\lambda=0$, symmetric splitting. (b) $\delta\lambda>0$ and (c) $\delta\lambda<0$: the doublet becomes asymmetric and the asymmetry sign flips with $\mathrm{sgn}(\delta\lambda)$. In the main text, neighboring longitudinal pairs sample different $\delta\lambda$ because of the Vernier mismatch between the TE$_0$ and TM$_0$ families.}
    \label{fig:tcmt_cases}
\end{figure}

Near degeneracy ($|\Delta|\ll g$, equivalently small $|\delta\lambda|$), the doublet splitting reduces to $\Omega=\omega_+-\omega_-\approx 2g$. Away from degeneracy, the doublet separation depends on both $g$ and the bare-mode detuning. Supplementary Fig.~\ref{fig:tcmt_cases} shows representative spectra computed from this model. The dashed vertical line marks the position of the dark mode. Defining $\delta\lambda \equiv \lambda_d-\lambda_b$, we see that at $\delta\lambda=0$ the doublet is symmetric and centered on the bare resonance. A finite $\delta\lambda$ unbalances the hybridization: one branch becomes more bright-like and the other more dark-like, and the doublet is correspondingly asymmetric. The roles of the two branches swap when $\delta\lambda$ changes sign.

\subsection{Photorefractive grating}

Under resonant pumping of the symmetric mode at \(\omega_p\), both mode families are populated inside the cavity and the pump field is written as
\begin{equation}
  \mathbf E_p^S(\mathbf r_\perp,s,t)
  =
  \left[
  a_b^{(p)}\mathbf e_b(\mathbf r_\perp,s;\omega_p)e^{i\beta_b(\omega_p)s}
  +
  a_d^{(p)}\mathbf e_d(\mathbf r_\perp,s;\omega_p)e^{i\beta_d(\omega_p)s}
  \right]e^{-i\omega_p t}.
  \label{eq:pump_field}
\end{equation}
Here \(s\) is the arclength along the racetrack and \(\mathbf e_{b,d}\) are the vectorial transverse mode profiles. We keep their possible \(s\)-dependence to account for the change of propagation direction, crystal-axis projection, and bend-induced mode mixing along the racetrack.

The photorefractive writing is driven by the local time-averaged optical intensity. Averaging the pump field over the optical period gives
\begin{align}
  I(\mathbf r_\perp,s)
  &\propto
  |a_b^{(p)}|^2|\mathbf e_b(\mathbf r_\perp,s;\omega_p)|^2
  +
  |a_d^{(p)}|^2|\mathbf e_d(\mathbf r_\perp,s;\omega_p)|^2
  \nonumber\\
  &\quad+
  2\,\mathrm{Re}\!\left[
  a_b^{(p)}a_d^{(p)\ast}
  \mathbf e_b(\mathbf r_\perp,s;\omega_p)\cdot
  \mathbf e_d^\ast(\mathbf r_\perp,s;\omega_p)
  e^{ik_G s}
  \right],
  \label{eq:pump_intensity}
\end{align}
where
\begin{equation}
  k_G=\beta_b(\omega_p)-\beta_d(\omega_p).
  \label{eq:kG}
\end{equation}
The first two terms are the individual traveling-wave modal intensities, which do not write a grating with wavevector \(k_G\). The last term is the intermodal interference term; it carries the longitudinal wavevector \(k_G\) and therefore writes the grating component that can phase-match coupling between the two co-propagating mode families.

For the TE--TM mode pair considered here, the local vector overlap \(\mathbf e_b\cdot\mathbf e_d^\ast\) is weak in the straight sections because the two modes are nearly orthogonally polarized. In the racetrack bends, the changing propagation direction, crystal-axis projection, and bend-induced polarization mixing enhance this overlap. The resulting photorefractive grating is therefore expected to be written predominantly in the bending regions.

Keeping only the grating component relevant for intermodal coupling, we write the induced permittivity perturbation phenomenologically as
\begin{equation}
  \Delta\varepsilon(\mathbf r_\perp,s,t)
  =
  a_b^{(p)}a_d^{(p)\ast}
  C(\mathbf r_\perp,s,t)
  e^{ik_G s}
  + \mathrm{c.c.},
  \label{eq:Deps}
\end{equation}
where \(C(\mathbf r_\perp,s,t)\) is chosen as a real, slowly varying photorefractive grating envelope. It absorbs the transverse pump-mode beating profile, the local photorefractive writing efficiency, the space-charge and electro-optic response, and any non-uniformity of the writing process along the racetrack. The addition of the complex conjugate ensures that \(\Delta\varepsilon\) is real. The phase of \(a_b^{(p)}a_d^{(p)\ast}\) sets the spatial position of the written grating.

In analogy to the symmetric superposition of bright and dark mode, under resonant pumping of the anti-symmetric mode the field propagating inside the racetrack will have a field distribution 
\begin{equation}
  \mathbf E_p^{AS}(\mathbf r_\perp,s,t)
  =
  \left[
  a_b^{(p)}\mathbf e_b(\mathbf r_\perp,s;\omega_p)e^{i\beta_b(\omega_p)s}
  -
  a_d^{(p)}\mathbf e_d(\mathbf r_\perp,s;\omega_p)e^{i\beta_d(\omega_p)s}
  \right]e^{-i\omega_p t}.
\end{equation}

We observe that switching the pump from one hybrid resonance to the other (symmetric to anti-symmetric) changes the relative phase between the bright and dark components by \(\pi\). Equivalently,
\begin{equation}
  a_b^{(p)}a_d^{(p)\ast}
  \rightarrow
  -a_b^{(p)}a_d^{(p)\ast}.
  \label{eq:gPR_phase_flip}
\end{equation}

Consequently, the induced permittivity perturbation is now 
\begin{equation}
  \Delta\varepsilon(\mathbf r_\perp,s,t)
  =
  -(a_b^{(p)}a_d^{(p)\ast}
  C(\mathbf r_\perp,s,t)
  e^{ik_G s}
  + \mathrm{c.c.}).
  \label{eq:Depsreverse}
\end{equation}
In real space, this corresponds to shifting the longitudinal beating pattern by half a grating period,
\begin{equation}
  e^{ik_G s}
  \rightarrow
  -e^{ik_G s}
  =
  e^{i(k_Gs+\pi)} .
  \label{eq:half_period_shift}
\end{equation}
Thus the two orthogonal hybrid modes write photorefractive gratings with the same spatial frequency \(k_G\) but opposite phase. Pumping them in sequence can therefore cancel or reverse the previously written contribution to \(g_{\mathrm{PR}}\), providing the mechanism for the write--erase--rewrite operation described in the main text.

\subsection{Coupling strength}

Treating \(\Delta\varepsilon\) as a weak perturbation, the photorefractive contribution to the intermodal coupling strength at probe frequency \(\omega\) is defined as:
\begin{equation}
  g_{\mathrm{PR}}(\omega,t)
  =
  \frac{\omega}{4\sqrt{U_b U_d}}
  \int_V
  \Delta\varepsilon(\mathbf r_\perp,s,t)\,
  \mathbf e_b(\mathbf r_\perp,s;\omega)
  \!\cdot\!
  \mathbf e_d^\ast(\mathbf r_\perp,s;\omega)\,
  e^{-i[\beta_b(\omega)-\beta_d(\omega)]s}\,dV,
  \label{eq:gPR_reciprocity}
\end{equation}

where \(U_{b,d}\) are the modal energies. Inserting Eq.~\eqref{eq:Deps} and retaining the phase-matched grating component gives
\begin{equation}
  g_{\mathrm{PR}}(\omega,t)
  =
  \frac{\omega}{4\sqrt{U_b U_d}}\,
  a_b^{(p)}\,a_d^{(p)*}\,
  \Gamma_{\mathrm{PR}}(\omega,t),
  \label{eq:gPR_general}
\end{equation}
with the generalized photorefractive overlap
\begin{equation}
  \Gamma_{\mathrm{PR}}(\omega,t)
  =
  \int_{-L/2}^{L/2} Q(s;\omega,t)e^{-i\Delta\beta(\omega)s}\,ds,
  \label{eq:GammaPR}
\end{equation}
where
\begin{equation}
  Q(s;\omega,t)
  =
  \int_A
  C(\mathbf r_\perp,s,t)\,
  \mathbf e_b(\mathbf r_\perp,s;\omega)
  \!\cdot\!
  \mathbf e_d^\ast(\mathbf r_\perp,s;\omega)\,dA.
  \label{eq:Q_profile}
\end{equation}
The residual phase mismatch is
\begin{equation}
  \Delta\beta(\omega)
  =
  [\beta_b(\omega)-\beta_d(\omega)]
  -
  [\beta_b(\omega_p)-\beta_d(\omega_p)].
  \label{eq:DeltaBeta}
\end{equation}
The profile \(Q(s;\omega,t)\) is the effective coupling profile of the written photorefractive grating. It contains both the non-uniform longitudinal writing strength and the local transverse coupling efficiency between the photorefractive-induced perturbation and the two optical modes.

For the build-up dynamics, we use the phenomenological single-time-constant form
\begin{equation}
  C(\mathbf r_\perp,s,t)
  =
  C_\infty(\mathbf r_\perp,s)
  \left(1-e^{-t/\tau_{\mathrm{PR}}}\right),
  \label{eq:C_build}
\end{equation}
where \(\tau_{\mathrm{PR}}\) is the photorefractive response time and \(C_\infty(\mathbf r_\perp,s)\) is the saturated grating envelope. This gives
\begin{equation}
  \Gamma_{\mathrm{PR}}(\omega,t)
  =
  \left(1-e^{-t/\tau_{\mathrm{PR}}}\right)
  \Gamma_{\mathrm{PR}}^{(\infty)}(\omega),
  \label{eq:GammaPR_build}
\end{equation}
with
\begin{equation}
  \Gamma_{\mathrm{PR}}^{(\infty)}(\omega)
  =
  \int_{-L/2}^{L/2} Q_\infty(s;\omega)e^{-i\Delta\beta(\omega)s}\,ds,
  \label{eq:GammaPR_inf}
\end{equation}
and
\begin{equation}
  Q_\infty(s;\omega)
  =
  \int_A
  C_\infty(\mathbf r_\perp,s)\,
  \mathbf e_b(\mathbf r_\perp,s;\omega)
  \!\cdot\!
  \mathbf e_d^\ast(\mathbf r_\perp,s;\omega)\,dA.
  \label{eq:Q_inf}
\end{equation}

At the writing frequency, \(\Delta\beta(\omega_p)=0\), and Eq.~\eqref{eq:gPR_general} reduces to
\begin{equation}
  g_{\mathrm{PR}}(\omega_p,t)
  =
  \eta_{\mathrm{eff}}
  \left(1-e^{-t/\tau_{\mathrm{PR}}}\right)
  a_b^{(p)}\,a_d^{(p)*},
  \label{eq:gPR_pump}
\end{equation}
where
\begin{equation}
  \eta_{\mathrm{eff}}
  =
  \frac{\omega_p}{4\sqrt{U_b U_d}}
  \int_{-L/2}^{L/2} Q_\infty(s;\omega_p)\,ds .
  \label{eq:eta_eff}
\end{equation}
This is the compact form quoted in the main text. The coefficient \(\eta_{\mathrm{eff}}\) is an effective writing efficiency that absorbs the saturated photorefractive response, the spatial distribution of the written grating, and the local intermodal coupling efficiency around the racetrack.

\subsection{Phase-matching bandwidth}

Away from the writing frequency, the frequency dependence of $g_{\mathrm{PR}}$ is governed mainly by the residual phase mismatch $\Delta\beta(\omega)$ in Eq.~\eqref{eq:DeltaBeta}. Over the spectral range considered here, we neglect the weaker frequency dependence of the local mode profiles and of the photorefractive perturbation itself, and evaluate the effective coupling profile at the writing frequency. We therefore define the normalized phase-matching function
\begin{equation}
  \Phi(\omega)
  =
  \frac{
  \int_{-L/2}^{L/2}
  Q_\infty(s;\omega_p)
  e^{-i\Delta\beta(\omega)s}\,ds
  }{
  \int_{-L/2}^{L/2}
  Q_\infty(s;\omega_p)\,ds
  }.
  \label{eq:Phi_general}
\end{equation}
The PR-induced coupling can then be written as
\begin{equation}
  g_{\mathrm{PR}}(\omega,t)
  \simeq
  g_{\mathrm{PR}}(\omega_p,t)\,\Phi(\omega),
  \label{eq:gPR_Phi}
\end{equation}
up to the slowly varying prefactor $\omega/\omega_p\simeq 1$.

To obtain a simple analytical estimate of Equation~\eqref{eq:Phi_general}, we approximate this profile as uniform over an effective interaction length $L_{\mathrm{eff}}$ and negligible outside this region:
\begin{equation}
  Q_\infty(s;\omega_p)
  \simeq
  Q_0,
  \qquad
  {-L_{\mathrm{eff}}/2}<s<{L_{\mathrm{eff}}/2}.
\end{equation}
Then
\begin{equation}
  \Phi(\omega)
  \simeq
  \frac{1}{L_{\mathrm{eff}}}
  \int_{-L_{\mathrm{eff}}/2}^{L_{\mathrm{eff}}/2}
  e^{-i\Delta\beta(\omega)s}\,ds .
\end{equation}
Evaluating the integral gives
\begin{equation}
  \Phi(\omega)
  =
  \mathrm{sinc}\!\left[
  \frac{\Delta\beta(\omega)L_{\mathrm{eff}}}{2}
  \right],
  \label{eq:Phi_sinc}
\end{equation}
where $\mathrm{sinc}(x)=\sin x/x$. Therefore
\begin{equation}
  g_{\mathrm{PR}}(\omega,t)
  \simeq
  g_{\mathrm{PR}}(\omega_p,t)\,
  \mathrm{sinc}\!\left[
  \frac{\Delta\beta(\omega)L_{\mathrm{eff}}}{2}
  \right].
  \label{eq:sinc_envelope}
\end{equation}
The bandwidth $\Delta\omega_{\mathrm{bw}}$ is defined as the FWHM of the coupling-strength envelope \(g_{\mathrm{PR}}(\omega,t)\). The condition
\begin{equation}
  \mathrm{sinc}(x_{\mathrm{bw}})=\frac{1}{2}
\end{equation}
has the solution $x_{\mathrm{bw}}\approx 1.8955$, so
\begin{equation}
  \Delta\omega_{\mathrm{bw}}
  \approx
  \frac{4c\,x_{\mathrm{bw}}}{|\Delta n_g|L_{\mathrm{eff}}}
  \approx
  \frac{7.58\,c}{|\Delta n_g|L_{\mathrm{eff}}},
  \qquad
  \Delta f_{\mathrm{bw}}=\frac{\Delta\omega_{\mathrm{bw}}}{2\pi}.
  \label{eq:bandwidth}
\end{equation}
Here $L_{\mathrm{eff}}$ could be understood as the coherent interaction length associated with the effective coupling profile $Q_\infty(s;\omega_p)$, rather than simply the geometric racetrack perimeter. A non-uniformly written grating, for example one that is stronger in the bends where TE0--TM0 beating and photorefractive writing are enhanced, is therefore captured by an effective length shorter than the full perimeter, as observed experimentally.

\section*{Supplementary Note 2. Device geometry}

\begin{figure}[!htbp]
    \centering
    \includegraphics[width=16 cm]{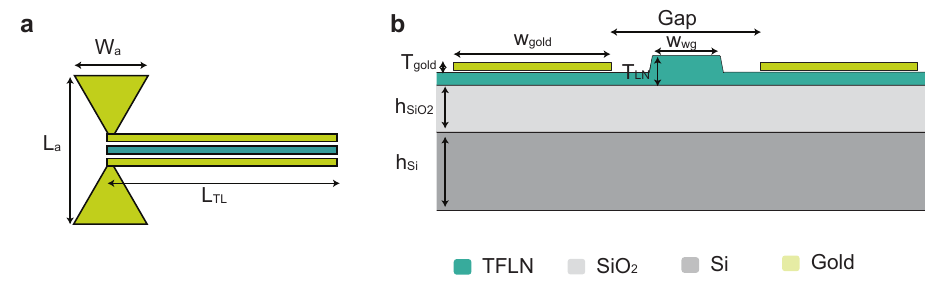}
    \caption{\textbf{Device geometry.}
    \textbf{a,} Top view of the coplanar transmission line with bowtie antenna.
    \textbf{b,} Cross-section of the lithium niobate rib waveguide and adjacent gold electrodes.
    Symbols correspond to the dimensions in Supplementary Table~\ref{tab:device_geometry}.}
    \label{fig:device_geometry}
\end{figure}

Supplementary Fig.~\ref{fig:device_geometry} shows the photonic and microwave geometry, and Supplementary Table~\ref{tab:device_geometry} lists the corresponding dimensions. The coplanar gold transmission line runs alongside the straight section of the racetrack. For the mmWave transduction device, a bowtie antenna is added at the end of the transmission line.

\begin{table}[!htbp]
\centering
\caption{\textbf{Device dimensions.}}
\label{tab:device_geometry}
\begin{tabular}{lll}
\hline
Parameter & Symbol & Value \\
\hline
Lithium niobate thickness & \(h_\mathrm{LN}\) & \(600~\mathrm{nm}\) \\
Buried oxide thickness & \(h_{\mathrm{SiO}_2}\) & \(4.7~\mu\mathrm{m}\) \\
Silicon substrate thickness & \(h_\mathrm{Si}\) & 500~\(\mu\mathrm{m}\) \\
Racetrack waveguide width & \(w_\mathrm{wg}\) & \(2~\mu\mathrm{m}\) \\
Gold thickness & \(T_\mathrm{gold}\) & \(300~\mathrm{nm}\) \\
Signal--ground gap & \(Gap\) & \(7~\mu\mathrm{m}\) \\
Gold strip width & \(w_\mathrm{gold}\) & 7~\(\mu\mathrm{m}\) \\
Transmission-line length & \(L_\mathrm{TL}\) & 750~\(\mu\mathrm{m}\) \\
Antenna width & \(W_\mathrm{a}\) & 300~\(\mu\mathrm{m}\) \\
Antenna length & \(L_\mathrm{a}\) & 407~\(\mu\mathrm{m}\) \\
\hline
\end{tabular}
\end{table}

\section*{Supplementary Note 3. Identification of bright and dark mode families}

We identify the bright and dark mode families by matching their measured group indices to COMSOL Multiphysics eigenmode simulations. On X-cut lithium niobate, both the effective and group indices depend on the in-plane angle $\theta$ between the propagation direction and the crystal axes ($\theta = 0$ corresponds to propagation along the Y-crystal direction). We therefore simulate the eigenmodes as a function of $\theta$, compute the angle-dependent group index
\begin{equation}
    n_g(\theta) = n_\mathrm{eff}(\theta)
    - \lambda \frac{\partial n_\mathrm{eff}(\theta)}{\partial \lambda},
    \label{eq:sim_ng_angle}
\end{equation}
and average it over the racetrack optical path,
\begin{equation}
    \overline{n}_g = \frac{1}{L}\int_0^L n_g[\theta(s)]\,ds,
    \label{eq:sim_ng_avg}
\end{equation}
where $L$ is the round-trip length and $\theta(s)$ the local propagation angle.

\begin{figure}[!t]
    \centering
    \includegraphics[width=16cm]{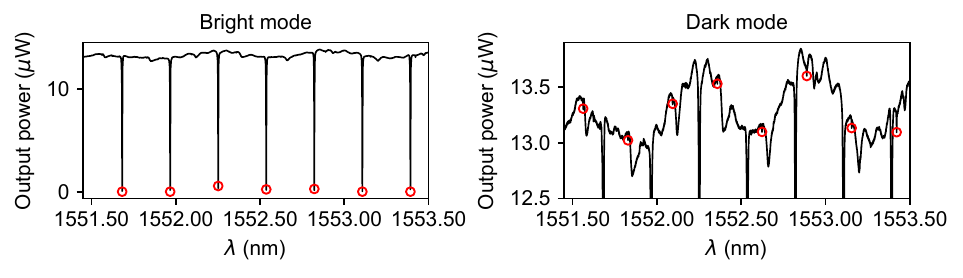}
    \caption{
    \textbf{Low-power transmission spectrum recorded before optical pumping.} Black curves: measured transmission of the bright (left) and dark (right) mode families. Red circles: resonances selected for the group-index extraction.
    }
    \label{fig:ng_resonances}
\end{figure}

\begin{figure}[!t]
    \centering
    \includegraphics[width=16cm]{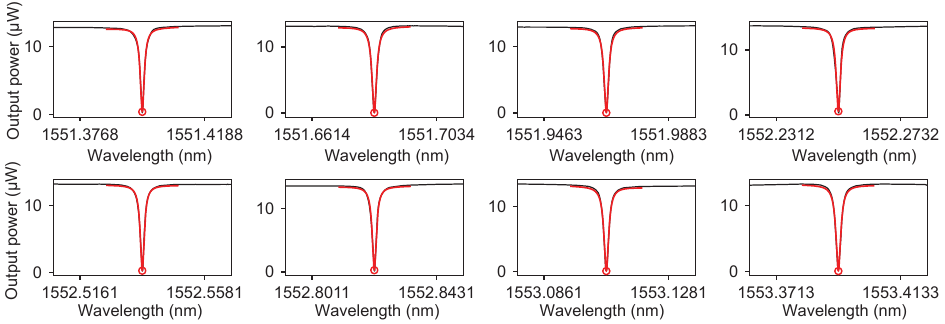}
    \caption{
    \textbf{Zoomed-in spectra of the eight bright-mode resonances used for the extraction of group index and quality factor.} Black curves: measured transmission spectra. Red curves: Lorentzian fits. Red circles: fitted resonance positions.
    }
    \label{fig:ng_bright_zoom}
\end{figure}

\begin{figure}[!t]
    \centering
    \includegraphics[width=16cm]{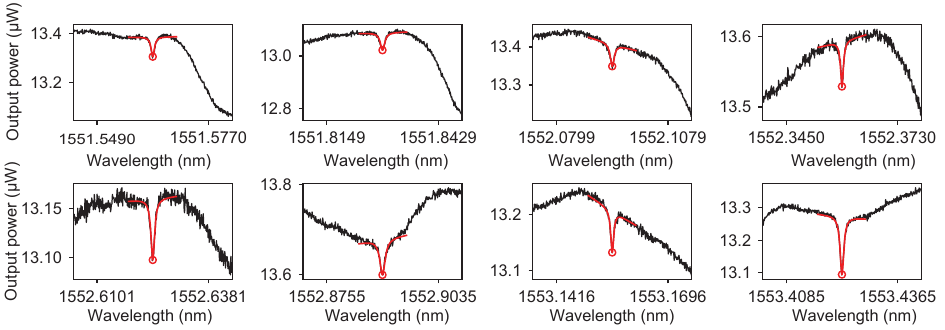}
    \caption{
    \textbf{Zoomed-in spectra of the eight dark-mode resonances used for the extraction of group index and quality factor.} Black curves: measured transmission spectra. Red curves: Lorentzian fits. Red circles: fitted resonance positions.
    }
    \label{fig:ng_dark_zoom}
\end{figure}

Experimentally, the group index of each family is extracted from the frequency-domain free spectral range, $n_g = c/(L\,\mathrm{FSR}_f)$, where $c$ is the speed of light in vacuum. To avoid pump-induced photorefractive coupling and mode hybridization, we use the transmission spectrum recorded before high-power optical pumping (Supplementary Fig.~\ref{fig:ng_resonances}). For each family, we select eight consecutive resonances and fit them with a Lorentzian lineshape (Supplementary Figs.~\ref{fig:ng_bright_zoom} and~\ref{fig:ng_dark_zoom}). The fitted resonance wavelengths $\lambda_m$ are converted to optical frequencies $f_m = c/\lambda_m$, and the local FSR is computed between adjacent resonances. The value reported in main text Fig.~2b is the mean of the resulting $n_g$ values, and the error bar is their standard deviation. The loaded quality factors obtained from these fits are shown in Supplementary Fig.~\ref{fig:ng_Q}.

In main text Fig.~2b, the measured group index of the dark family falls between the simulated values for TM$_0$ and TE$_2$, so the assignment cannot be made from $n_g$ alone. However, since TE$_2$ is less transversely confined than TE$_0$ and TM$_0$, its field overlaps more strongly with the adjacent gold electrodes, placing a limit on their quality factors. The simulated absorption-limited intrinsic $Q$ for TE$_2$ is $\sim 3\times 10^5$, well below the measured loaded $Q$ of the dark resonances (Supplementary Fig.~\ref{fig:ng_Q}). We therefore rule out TE$_2$ and assign the dark family to TM$_0$.

\begin{figure}[!t]
    \centering
    \includegraphics[width=8cm]{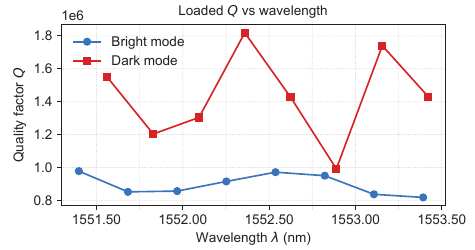}
    \caption{
    \textbf{Loaded quality factors of the bright and dark mode families,} extracted from the Lorentzian fits in Supplementary Figs.~\ref{fig:ng_bright_zoom} and~\ref{fig:ng_dark_zoom}.
    }
    \label{fig:ng_Q}
\end{figure}

\section*{Supplementary Note 4. Analysis of hybrid-mode spectra}

We determine the intermodal coupling strength $g$ by fitting the measured hybrid doublets with the steady-state two-mode transmission function in Eq.~\eqref{eq:transmission}. The bus-coupled resonance is assigned to the TE$_0$ mode family, while the weakly excited resonance is assigned to the TM$_0$ mode family, following the modal identification described in Supplementary Note~3.

Before fitting, each spectrum is normalized to the local off-resonant transmission. The resonance positions are initialized from the measured transmission minima in the fitting window. The fit returns the bare TE$_0$ and TM$_0$ resonance wavelengths, the intermodal coupling strength $g$, the linewidths of the two modes, and the external coupling rate of the bright mode. Representative fits are shown in Supplementary Fig.~\ref{fig:representative_tcmt_fits}. The two-mode model provides good agreement with the measured spectra over the different coupling regimes, as shown by the representative fits in Supplementary Fig.~\ref{fig:representative_tcmt_fits}.

\begin{figure}[t]
    \centering
    \includegraphics[width= 16 cm]{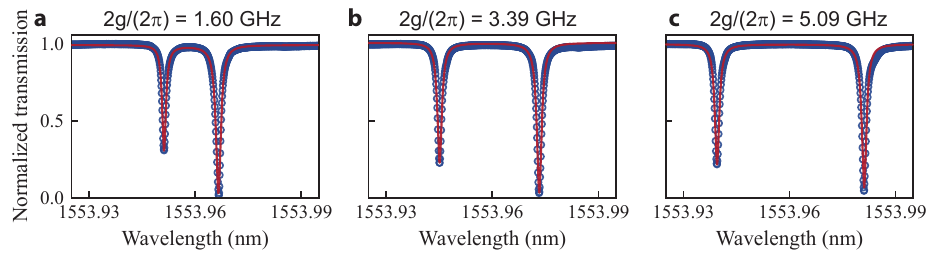}
    \caption{
    Representative fits of the hybrid doublets. Blue markers are measured
    normalized transmission spectra, and red curves are fits using the two-mode
    transmission function in Eq.~\eqref{eq:transmission}. The extracted
    splittings are (a) $2g/2\pi=1.60~\mathrm{GHz}$,
    (b) $2g/2\pi=3.39~\mathrm{GHz}$, and
    (c) $2g/2\pi=5.09~\mathrm{GHz}$.
    }
    \label{fig:representative_tcmt_fits}
\end{figure}

For the power-dependent writing measurements, this fitting procedure is applied
to every recorded spectrum to obtain $g$ as a function of iteration number and
on-chip pump power. For the bandwidth measurement, the coupling strengths
extracted from different resonance pairs are fitted with the phase-matching
envelope derived in Supplementary Note~1,
\begin{equation}
    2g(f) =
    2g_0\,
    \mathrm{sinc}
    \left[
    \frac{\Delta n_g L_{\mathrm{eff}}}{c}\pi(f-f_p)
    \right],
    \label{eq:sinc_fit}
\end{equation}
where $f_p$ is the writing frequency, $\Delta n_g$ is the TE$_0$--TM$_0$
group-index difference, and $L_{\mathrm{eff}}$ is the effective interaction
length of the written grating. The bandwidth is taken as the FWHM of this fitted
envelope.

The same fitting routine is used for the relaxation measurement after the pump
is switched off. The extracted $g(t)$ is then fitted with the phenomenological
decay model used in the main text.

\section*{Supplementary Note 5. Optical programming of the coupling strength by write power}

\label{sec:power_dependent_coupling}
\begin{figure}[!tbp]
  \centering
  \includegraphics[width=16cm]{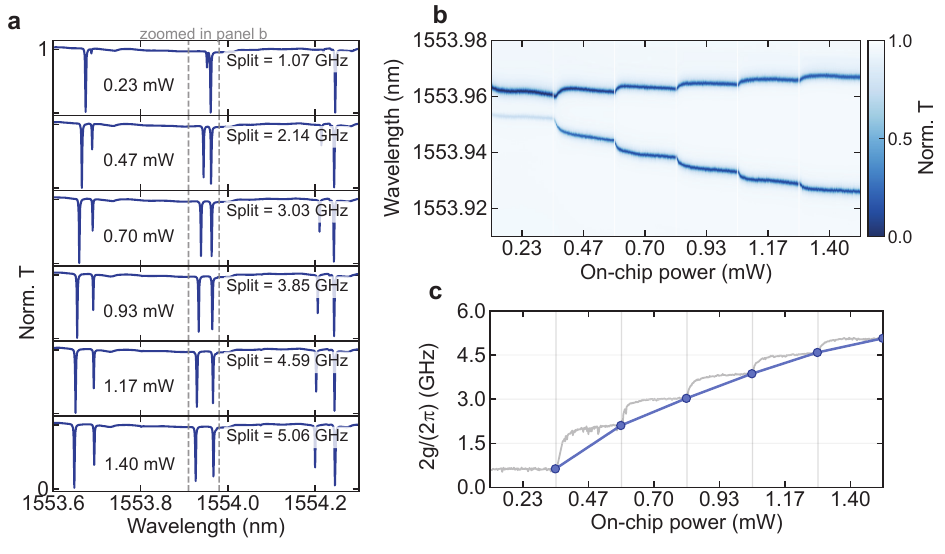}
  \caption{\textbf{Optical programming of the photorefraction-induced intermodal coupling.}
  (a) Normalized transmission spectra recorded after saturation of the photorefractive grating for different estimated on-chip write powers in the bus waveguide.
  The progressive increase of the hybrid-mode splitting indicates stronger intermodal coupling.
  (b) Evolution of the transmission spectrum during the iterative write--probe sequence.
  The estimated on-chip write power is increased from 0.233~mW to 1.398~mW in steps of 0.233~mW, and the device is pumped for 100 iterations at each power setting.
  (c) Extracted intermodal coupling strength $2g/2\pi$ as a function of iteration number and write power.
  Near zero bright--dark detuning, $2g/2\pi$ corresponds to the observable hybrid-mode splitting.
  Gray traces show the extracted $2g/2\pi$ during the writing dynamics, while blue markers indicate the saturated coupling strength for each write power.}
  \label{fig:power_dependence}
\end{figure}

To verify that the photorefraction-induced intermodal coupling can be optically programmed, we measured the evolution of the hybrid-mode splitting as a function of the write power.
During the writing process, we used the same iterative write--probe protocol as described in the Methods.
Each iteration consists of a low-power wavelength sweep to locate the instantaneous resonance position, a resonant write step at the selected wavelength, and a second low-power probe sweep to record the updated transmission spectrum.
The iteration number therefore provides a practical measure of the cumulative on-resonance optical exposure applied to the device.

In this measurement, the estimated on-chip write power in the bus waveguide was increased from 0.233~mW to 1.398~mW in steps of 0.233~mW.
For each power setting, the device was pumped for 100 iterations, corresponding to a total duration of approximately 1.38~h.
Representative transmission spectra recorded after the coupling strength reached saturation are shown in Fig.~\ref{fig:power_dependence}a.
As the write power is increased, the saturated hybrid-mode splitting becomes progressively larger, indicating an increase of the effective intermodal coupling strength.
The full spectral evolution of the pumped hybrid-mode pair is shown in Fig.~\ref{fig:power_dependence}b.
For each fixed write power, the splitting first grows rapidly over the initial iterations and then approaches a steady-state value, consistent with saturation of the written photorefractive grating.

To quantify the coupling dynamics, we fitted each measured hybrid doublet using the two-mode transmission model described in Supplementary Note 1.
The extracted coupling strengths are summarized in Fig.~\ref{fig:power_dependence}c.
Close to local degeneracy between the bright and dark modes, the hybrid-mode splitting is approximately equal to $2g$. 
Therefore, the fitted $2g$ values can be interpreted directly as the observable doublet splitting.
The data show that $2g$ increases during the writing process and saturates for each write power.
The saturated coupling strength also increases monotonically with the applied write power, demonstrating that the final value of the photorefraction-induced coupling can be set optically.

The dependence of the saturated coupling strength on write power is sublinear over the measured range.
In particular, the incremental increase of $2g$ becomes smaller at higher write powers.
This behavior is consistent with saturation of the photorefractive writing process.
Possible microscopic contributions include trap filling, redistribution of space-charge fields, and screening of the internal electric field at higher optical intensities~\cite{Jermann:93,PhysRevB.78.245114}.
Together, these measurements establish that the photorefractive grating provides an optically programmable contribution to the intermodal coupling strength, which is used in the main text to tune the mmWave transduction frequency.

\section*{Supplementary Note 6. Temperature tuning of the bright--dark detuning}

\begin{figure}[t]
    \centering
    \includegraphics[width=16cm]{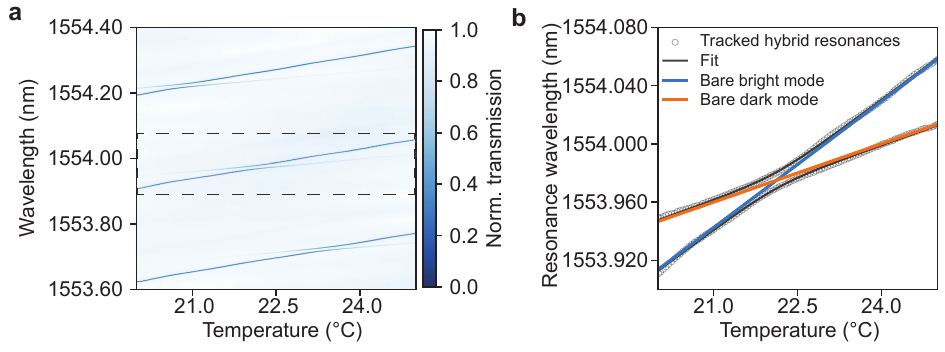}
    \caption{
    \textbf{Temperature tuning of the photonic molecule.}
    \textbf{a,} Transmission spectra recorded after photorefractive grating writing as the chip is swept from 20 to 25~$^\circ$C. The bright and dark families shift at different rates with temperature, so the bright--dark detuning of every longitudinal pair is temperature-dependent. The dashed box marks the pair analyzed in panel \textbf{b}.
    \textbf{b,} Tracked hybrid resonances of the boxed pair (markers) and two-mode fit using Eq.~\eqref{eq:temperature_two_mode} (black). Bare bright and dark modes from the fit are shown in blue and orange.
    }
    \label{fig:temperature_tuning}
\end{figure}

The hybrid spectrum of the photonic molecule is set by two parameters: the intermodal coupling $g$ and the bare-mode detuning. Optical writing programs $g$ (main text Fig.~3); temperature shifts the bare-mode detuning, and therefore selects which longitudinal pair sits closest to degeneracy.

We wrote a photorefractive grating on one hybrid resonance and then swept the chip temperature from 20 to 25~$^\circ$C in 50~mK steps on a thermoelectric mount (Thorlabs PTC1/M). At each step we recorded a low-power probe spectrum to avoid disturbing the grating. All hybrid pairs red-shift with temperature, but the bright and dark families do so at different rates, so each pair sweeps through a temperature-dependent detuning (Supplementary Fig.~\ref{fig:temperature_tuning}a). We focus on a single pair (dashed box) whose detuning crosses zero in the swept range.

Assuming $g$ is independent of temperature over the 5~K range and that both bare resonances tune linearly, $\lambda_{b,d}(T) = \lambda_{b,d}(T_0) + \alpha_{b,d}(T-T_0)$, the hybrid resonances follow
\begin{equation}
    \lambda_{\pm}(T)
    = \tfrac{1}{2}\bigl[\lambda_b(T)+\lambda_d(T)\bigr]
    \pm \sqrt{\tfrac{1}{4}\bigl[\lambda_b(T)-\lambda_d(T)\bigr]^2 + g^2},
    \label{eq:temperature_two_mode}
\end{equation}
with $g$ in wavelength units. Fitting the tracked resonances (Supplementary Fig.~\ref{fig:temperature_tuning}b) gives $\alpha_b = 28.9$~pm/K, $\alpha_d = 13.3$~pm/K, and $2g/2\pi \approx 1.4$~GHz at $\lambda_p = 1554$~nm. The faster tuning of the bright mode reflects the anisotropic thermo-optic response of x-cut TFLN: the effective coefficient depends on both polarization and propagation direction relative to the crystal axes~\cite{Shim2026ThermoOpticLN}. The TE$_0$ field projects mostly onto the extraordinary axis, which has the larger thermo-optic coefficient, while TM$_0$ samples almost exclusively the ordinary axis.

The fact that a single value of $g$ describes the full scan confirms that temperature affects only the bare-mode detuning. Together with optical writing of $g$, this gives two independent post-fabrication knobs: the photorefractive grating sets the coupling strength, and temperature places a chosen longitudinal pair at the degeneracy point with the most efficient photorefractive grating writing.

\newpage
\section*{References}
\bibliographystyle{paper}
\bibliography{bibliography}

@article{Shim2026ThermoOpticLN,
  author  = {Shim, Joonsup and Kim, Seonghun and Lu, Shengyuan and Yang, Jiayu and Jeon, Seongjin and Kim, SangHyeon and Lon{\v{c}}ar, Marko and Sohn, Young-Ik},
  title   = {Generalized Model of Anisotropic Thermo-Optic Response on Thin-Film Lithium Niobate Platform},
  journal = {ACS Photonics},
  volume  = {13},
  number  = {9},
  pages   = {2586--2596},
  year    = {2026},
  doi     = {10.1021/acsphotonics.6c00117}
}

@article{Jermann:93,
author = {F. Jermann and J. Otten},
journal = {J. Opt. Soc. Am. B},
number = {11},
pages = {2085--2092},
publisher = {Optica Publishing Group},
title = {Light-induced charge transport in LiNbO3:Fe at high light intensities},
volume = {10},
month = {Nov},
year = {1993},
url = {https://opg.optica.org/josab/abstract.cfm?URI=josab-10-11-2085},
doi = {10.1364/JOSAB.10.002085}
}

@article{PhysRevB.78.245114,
  title = {Light-induced charge transport in ${\text{LiNbO}}_{3}$ crystals},
  author = {Sturman, B. and Carrascosa, M. and Agullo-Lopez, F.},
  journal = {Phys. Rev. B},
  volume = {78},
  issue = {24},
  pages = {245114},
  numpages = {8},
  year = {2008},
  month = {Dec},
  publisher = {American Physical Society},
  doi = {10.1103/PhysRevB.78.245114},
  url = {https://link.aps.org/doi/10.1103/PhysRevB.78.245114}
}

@article{PhysRevLett.127.033902,
  title = {Resonant Stimulated Photorefractive Scattering},
  author = {Liu, Jingliang and Stace, Thomas and Dai, Jian and Xu, Kun and Luiten, Andre and Baynes, Fred},
  journal = {Phys. Rev. Lett.},
  volume = {127},
  issue = {3},
  pages = {033902},
  numpages = {6},
  year = {2021},
  month = {Jul},
  publisher = {American Physical Society},
  doi = {10.1103/PhysRevLett.127.033902},
  url = {https://link.aps.org/doi/10.1103/PhysRevLett.127.033902}
}

@article{gaier2025,
  title   = {Wireless millimeter-wave electro-optics on thin-film lithium niobate},
  author  = {Gaier, A. and Mamian, K. and Rajabali, S. and Lampert, Y. and Liu, J. and Magalhaes, L. and Shams-Ansari, A. and Loncar, M. and Benea-Chelmus, I.-C.},
  journal = {arXiv:2505.04585},
  year    = {2025}
}

@article{PhysRevA.90.053841,
  title = {Electromagnetically-induced-transparency-like ground-state cooling in a double-cavity optomechanical system},
  author = {Guo, Yujie and Li, Kai and Nie, Wenjie and Li, Yong},
  journal = {Phys. Rev. A},
  volume = {90},
  issue = {5},
  pages = {053841},
  numpages = {6},
  year = {2014},
  month = {Nov},
  publisher = {American Physical Society},
  doi = {10.1103/PhysRevA.90.053841},
  url = {https://link.aps.org/doi/10.1103/PhysRevA.90.053841}
}

@article{SantamariaBotello:18,
author = {Gabriel Santamar{\'i}a Botello and Florian Sedlmeir and Alfredo Rueda and Kerlos Atia Abdalmalak and Elliott R. Brown and Gerd Leuchs and Sascha Preu and Daniel Segovia-Vargas and Dmitry V. Strekalov and Luis Enrique Garc{\'i}a Mu{\~n}oz and Harald G. L. Schwefel},
journal = {Optica},
number = {10},
pages = {1210--1219},
publisher = {Optica Publishing Group},
title = {Sensitivity limits of millimeter-wave photonic radiometers based on efficient electro-optic upconverters},
volume = {5},
month = {Oct},
year = {2018},
url = {https://opg.optica.org/optica/abstract.cfm?URI=optica-5-10-1210},
doi = {10.1364/OPTICA.5.001210},
}

@article{hu_-chip_2021,
	title = {On-chip electro-optic frequency shifters and beam splitters},
	volume = {599},
	issn = {1476-4687},
	url = {https://doi.org/10.1038/s41586-021-03999-x},
	doi = {10.1038/s41586-021-03999-x},
	number = {7886},
	journal = {Nature},
	author = {Hu, Yaowen and Yu, Mengjie and Zhu, Di and Sinclair, Neil and Shams-Ansari, Amirhassan and Shao, Linbo and Holzgrafe, Jeffrey and Puma, Eric and Zhang, Mian and Lončar, Marko},
	month = nov,
	year = {2021},
	pages = {587--593},
}

@article{helgason_surpassing_2023,
	title = {Surpassing the nonlinear conversion efficiency of soliton microcombs},
	volume = {17},
	issn = {1749-4893},
	url = {https://doi.org/10.1038/s41566-023-01280-3},
	doi = {10.1038/s41566-023-01280-3},
	number = {11},
	journal = {Nature Photonics},
	author = {Helgason, {\'O}skar B. and Girardi, Marcello and Ye, Zhichao and Lei, Fuchuan and Schr{\"o}der, Jochen and Torres-Company, Victor},
	month = nov,
	year = {2023},
	pages = {992--999},
}

@article{zhou_self-organized_2025,
	title = {Self-organized spatiotemporal quasi-phase-matching in microresonators},
	volume = {16},
	issn = {2041-1723},
	url = {https://doi.org/10.1038/s41467-025-59215-1},
	doi = {10.1038/s41467-025-59215-1},
	number = {1},
	journal = {Nature Communications},
	author = {Zhou, Ji and Hu, Jianqi and Clementi, Marco and Yakar, Ozan and Nitiss, Edgars and Stroganov, Anton and Brès, Camille-Sophie},
	month = may,
	year = {2025},
	pages = {4083},
}

@article{nitiss_optically_2022,
	title = {Optically reconfigurable quasi-phase-matching in silicon nitride microresonators},
	volume = {16},
	issn = {1749-4893},
	url = {https://doi.org/10.1038/s41566-021-00925-5},
	doi = {10.1038/s41566-021-00925-5},
	number = {2},
	journal = {Nature Photonics},
	author = {Nitiss, Edgars and Hu, Jianqi and Stroganov, Anton and Brès, Camille-Sophie},
	month = feb,
	year = {2022},
	pages = {134--141},
}

@article{li2026polarization,
  title   = {One-dimensional polarization-hybrid photonic crystal molecules},
  author  = {Li, T. and Gallo, K.},
  journal = {arXiv:2605.03899},
  year    = {2026}
}

@article{li_down-converted_2025,
	title = {Down-converted photon pairs in a high-{Q} silicon nitride microresonator},
	volume = {639},
	issn = {1476-4687},
	url = {https://doi.org/10.1038/s41586-025-08662-3},
	doi = {10.1038/s41586-025-08662-3},
	number = {8056},
	journal = {Nature},
	author = {Li, Bohan and Yuan, Zhiquan and Williams, James and Jin, Warren and Beckert, Adrian and Xie, Tian and Guo, Joel and Feshali, Avi and Paniccia, Mario and Faraon, Andrei and Bowers, John and Marandi, Alireza and Vahala, Kerry},
	month = mar,
	year = {2025},
	pages = {922--927},
}

@article{aharonovich_programmable_2026,
	title = {Programmable integrated quantum photonics},
	volume = {20},
	issn = {1749-4893},
	url = {https://doi.org/10.1038/s41566-025-01830-x},
	doi = {10.1038/s41566-025-01830-x},
	number = {3},
	journal = {Nature Photonics},
	author = {Aharonovich, Igor and Crozier, Kenneth B. and Neshev, Dragomir},
	month = mar,
	year = {2026},
	pages = {254--265},
}

@article{Siliconmicroring,
author = {Bogaerts, W. and De Heyn, P. and Van Vaerenbergh, T. and De Vos, K. and Kumar Selvaraja, S. and Claes, T. and Dumon, P. and Bienstman, P. and Van Thourhout, D. and Baets, R.},
title = {Silicon microring resonators},
journal = {Laser \& Photonics Reviews},
volume = {6},
number = {1},
pages = {47-73},
doi = {https://doi.org/10.1002/lpor.201100017},
url = {https://onlinelibrary.wiley.com/doi/abs/10.1002/lpor.201100017},
eprint = {https://onlinelibrary.wiley.com/doi/pdf/10.1002/lpor.201100017},
year = {2012}
}

@article{bogaerts_programmable_2020,
	title = {Programmable photonic circuits},
	volume = {586},
	issn = {1476-4687},
	url = {https://doi.org/10.1038/s41586-020-2764-0},
	doi = {10.1038/s41586-020-2764-0},
	number = {7828},
	journal = {Nature},
	author = {Bogaerts, Wim and Pérez, Daniel and Capmany, José and Miller, David A. B. and Poon, Joyce and Englund, Dirk and Morichetti, Francesco and Melloni, Andrea},
	month = oct,
	year = {2020},
	pages = {207--216},
}

@article{perez-lopez_multipurpose_2020,
	title = {Multipurpose self-configuration of programmable photonic circuits},
	volume = {11},
	issn = {2041-1723},
	url = {https://doi.org/10.1038/s41467-020-19608-w},
	doi = {10.1038/s41467-020-19608-w},
	number = {1},
	journal = {Nature Communications},
	author = {Pérez-López, Daniel and López, Aitor and DasMahapatra, Prometheus and Capmany, José},
	month = dec,
	year = {2020},
	pages = {6359},
}

@article{churaev_heterogeneously_2023,
	title = {A heterogeneously integrated lithium niobate-on-silicon nitride photonic platform},
	volume = {14},
	issn = {2041-1723},
	url = {https://doi.org/10.1038/s41467-023-39047-7},
	doi = {10.1038/s41467-023-39047-7},
	number = {1},
	journal = {Nature Communications},
	author = {Churaev, Mikhail and Wang, Rui Ning and Riedhauser, Annina and Snigirev, Viacheslav and Blésin, Terence and Möhl, Charles and Anderson, Miles H. and Siddharth, Anat and Popoff, Youri and Drechsler, Ute and Caimi, Daniele and Hönl, Simon and Riemensberger, Johann and Liu, Junqiu and Seidler, Paul and Kippenberg, Tobias J.},
	month = jun,
	year = {2023},
	pages = {3499},
}

@article{li_high_2023,
	title = {High density lithium niobate photonic integrated circuits},
	volume = {14},
	issn = {2041-1723},
	url = {https://doi.org/10.1038/s41467-023-40502-8},
	doi = {10.1038/s41467-023-40502-8},
	number = {1},
	journal = {Nature Communications},
	author = {Li, Zihan and Wang, Rui Ning and Lihachev, Grigory and Zhang, Junyin and Tan, Zelin and Churaev, Mikhail and Kuznetsov, Nikolai and Siddharth, Anat and Bereyhi, Mohammad J. and Riemensberger, Johann and Kippenberg, Tobias J.},
	month = aug,
	year = {2023},
	pages = {4856},
}

@article{shekhar_roadmapping_2024,
	title = {Roadmapping the next generation of silicon photonics},
	volume = {15},
	issn = {2041-1723},
	url = {https://doi.org/10.1038/s41467-024-44750-0},
	doi = {10.1038/s41467-024-44750-0},
	number = {1},
	journal = {Nature Communications},
	author = {Shekhar, Sudip and Bogaerts, Wim and Chrostowski, Lukas and Bowers, John E. and Hochberg, Michael and Soref, Richard and Shastri, Bhavin J.},
	month = jan,
	year = {2024},
	pages = {751},
}

@article{Xue:22,
author = {Yu Xue and Ranfeng Gan and Kaixuan Chen and Gengxin Chen and Ziliang Ruan and Junwei Zhang and Jie Liu and Daoxin Dai and Changjian Guo and Liu Liu},
journal = {Optica},
number = {10},
pages = {1131--1137},
publisher = {Optica Publishing Group},
title = {Breaking the bandwidth limit of a high-quality-factor ring modulator based on thin-film lithium niobate},
volume = {9},
month = {Oct},
year = {2022},
url = {https://opg.optica.org/optica/abstract.cfm?URI=optica-9-10-1131},
doi = {10.1364/OPTICA.470596},
}

@article{mckenna_cryogenic_2020,
	title = {Cryogenic microwave-to-optical conversion using a triply resonant lithium-niobate-on-sapphire transducer},
	volume = {7},
	url = {https://opg.optica.org/optica/abstract.cfm?URI=optica-7-12-1737},
	doi = {10.1364/OPTICA.397235},
	number = {12},
	journal = {Optica},
	publisher = {Optica Publishing Group},
	author = {McKenna, Timothy P. and Witmer, Jeremy D. and Patel, Rishi N. and Jiang, Wentao and Laer, Raphaël Van and Arrangoiz-Arriola, Patricio and Wollack, E. Alex and Herrmann, Jason F. and Safavi-Naeini, Amir H.},
	month = dec,
	year = {2020},
	pages = {1737--1745},
}

@article{Multani2026,
	title = {Integrated millimeter-wave cavity electro-optic transduction},
	volume = {17},
	issn = {2041-1723},
	url = {https://doi.org/10.1038/s41467-025-67932-w},
	doi = {10.1038/s41467-025-67932-w},
	number = {1},
	journal = {Nature Communications},
	author = {Multani, Kevin K. S. and Herrmann, Jason F. and Nanni, Emilio A. and Safavi-Naeini, Amir H.},
	month = jan,
	year = {2026},
	pages = {1166},
}

@article{hsu_-chip_2023,
	title = {On-chip wavelength division multiplexing filters using extremely efficient gate-driven silicon microring resonator array},
	volume = {13},
	issn = {2045-2322},
	url = {https://doi.org/10.1038/s41598-023-32313-0},
	doi = {10.1038/s41598-023-32313-0},
	number = {1},
	journal = {Scientific Reports},
	author = {Hsu, Wei-Che and Nujhat, Nabila and Kupp, Benjamin and Conley, John F. and Wang, Alan X.},
	month = mar,
	year = {2023},
	pages = {5269},
}

@article{hu_high-efficiency_2022,
	title = {High-efficiency and broadband on-chip electro-optic frequency comb generators},
	volume = {16},
	issn = {1749-4893},
	url = {https://doi.org/10.1038/s41566-022-01059-y},
	doi = {10.1038/s41566-022-01059-y},
	number = {10},
	journal = {Nature Photonics},
	author = {Hu, Yaowen and Yu, Mengjie and Buscaino, Brandon and Sinclair, Neil and Zhu, Di and Cheng, Rebecca and Shams-Ansari, Amirhassan and Shao, Linbo and Zhang, Mian and Kahn, Joseph M. and Lončar, Marko},
	month = oct,
	year = {2022},
	pages = {679--685},
}

@article{song_hybrid_2025,
	title = {Hybrid {Kerr}-electro-optic frequency combs on thin-film lithium niobate},
	volume = {14},
	issn = {2047-7538},
	url = {https://doi.org/10.1038/s41377-025-01906-x},
	doi = {10.1038/s41377-025-01906-x},
	number = {1},
	journal = {Light: Science \& Applications},
	author = {Song, Yunxiang and Hu, Yaowen and Lončar, Marko and Yang, Kiyoul},
	month = aug,
	year = {2025},
	pages = {270},
}

@article{Tao2024VersatilePM,
  author  = {Tao, Zihan and Shen, Bitao and Li, Wencan and Xing, Luwen and Wang, Haoyu and Wu, Yichen and Tao, Yuansheng and Zhou, Yan and He, Yandong and Peng, Chao and Shu, Haowen and Wang, Xingjun},
  title   = {Versatile photonic molecule switch in multimode microresonators},
  journal = {Light: Science \& Applications},
  year    = {2024},
  volume  = {13},
  number  = {1},
  pages   = {51},
  doi     = {10.1038/s41377-024-01399-0}
}

@article{Shen2020ReconfigurableSplitting,
  author  = {Shen, Bin and Lin, Hongtao and Sharif Azadeh, Saeed and Nojic, Jovana and Kang, Myungkoo and Merget, Florian and Richardson, Kathleen A. and Hu, Juejun and Witzens, Jeremy},
  title   = {Reconfigurable Frequency-Selective Resonance Splitting in Chalcogenide Microring Resonators},
  journal = {ACS Photonics},
  year    = {2020},
  volume  = {7},
  number  = {2},
  pages   = {499--511},
  doi     = {10.1021/acsphotonics.9b01593}
}

@article{Xia2025ReconfigurableChG,
  author  = {Xia, Di and Luo, Liyang and Wang, Linyi and Zhao, Xin and Yang, Zelin and Wu, Jiayue and Yang, Qi-Fan and Li, Zhaohui and Zhang, Bin},
  title   = {Reconfigurable chalcogenide integrated nonlinear photonics},
  journal = {Nature Communications},
  year    = {2025},
  volume  = {16},
  pages   = {10133},
  doi     = {10.1038/s41467-025-65009-2}
}

@article{PhysRevA.96.043808,
  title = {Efficient quantum microwave-to-optical conversion using electro-optic nanophotonic coupled resonators},
  author = {Soltani, Mohammad and Zhang, Mian and Ryan, Colm and Ribeill, Guilhem J. and Wang, Cheng and Loncar, Marko},
  journal = {Phys. Rev. A},
  volume = {96},
  issue = {4},
  pages = {043808},
  numpages = {9},
  year = {2017},
  month = {Oct},
  publisher = {American Physical Society},
  doi = {10.1103/PhysRevA.96.043808},
  url = {https://link.aps.org/doi/10.1103/PhysRevA.96.043808}
}

@article{Peng:12,
author = {Bo Peng and \c{S}ahin Kaya \"{O}zdemir and Jiangang Zhu and Lan Yang},
journal = {Opt. Lett.},
number = {16},
pages = {3435--3437},
publisher = {Optica Publishing Group},
title = {Photonic molecules formed by coupled hybrid resonators},
volume = {37},
month = {Aug},
year = {2012},
url = {https://opg.optica.org/ol/abstract.cfm?URI=ol-37-16-3435},
doi = {10.1364/OL.37.003435},
}

@article{Zhang2019ProgrammablePM,
  author  = {Zhang, Mian and Wang, Cheng and Hu, Yaowen and Shams-Ansari, Amirhassan and Ren, Tianhao and Fan, Shanhui and Lon{\v{c}}ar, Marko},
  title   = {Electronically programmable photonic molecule},
  journal = {Nature Photonics},
  year    = {2019},
  volume  = {13},
  number  = {1},
  pages   = {36--40},
  doi     = {10.1038/s41566-018-0317-y},
  url     = {https://doi.org/10.1038/s41566-018-0317-y}
}

@article{Liu:20,
author = {Xiaoyue Liu and Pan Ying and Xuming Zhong and Jian Xu and Ya Han and Siyuan Yu and Xinlun Cai},
journal = {Opt. Lett.},
number = {22},
pages = {6318--6321},
publisher = {Optica Publishing Group},
title = {Highly efficient thermo-optic tunable micro-ring resonator based on an LNOI platform},
volume = {45},
month = {Nov},
year = {2020},
url = {https://opg.optica.org/ol/abstract.cfm?URI=ol-45-22-6318},
doi = {10.1364/OL.410192},
}

@article{Guarino2007,
	title = {Electro–optically tunable microring resonators in lithium niobate},
	volume = {1},
	issn = {1749-4893},
	url = {https://doi.org/10.1038/nphoton.2007.93},
	doi = {10.1038/nphoton.2007.93},
	number = {7},
	journal = {Nature Photonics},
	author = {Guarino, Andrea and Poberaj, Gorazd and Rezzonico, Daniele and Degl'Innocenti, Riccardo and Günter, Peter},
	month = jul,
	year = {2007},
	pages = {407--410},
}

@article{rakovich2010photonic,
  title={Photonic atoms and molecules},
  author={Rakovich, Yury P and Donegan, John F},
  journal={Laser \& Photonics Reviews},
  volume={4},
  number={2},
  pages={179--191},
  year={2010},
  publisher={Wiley Online Library}
}

@article{PhysRevLett.81.2582,
  title = {Optical Modes in Photonic Molecules},
  author = {Bayer, M. and Gutbrod, T. and Reithmaier, J. P. and Forchel, A. and Reinecke, T. L. and Knipp, P. A. and Dremin, A. A. and Kulakovskii, V. D.},
  journal = {Phys. Rev. Lett.},
  volume = {81},
  issue = {12},
  pages = {2582--2585},
  numpages = {0},
  year = {1998},
  month = {Sep},
  publisher = {American Physical Society},
  doi = {10.1103/PhysRevLett.81.2582},
  url = {https://link.aps.org/doi/10.1103/PhysRevLett.81.2582}
}

@article{Zhu:21,
author = {Di Zhu and Linbo Shao and Mengjie Yu and Rebecca Cheng and Boris Desiatov and C. J. Xin and Yaowen Hu and Jeffrey Holzgrafe and Soumya Ghosh and Amirhassan Shams-Ansari and Eric Puma and Neil Sinclair and Christian Reimer and Mian Zhang and Marko Lon\v{c}ar},
journal = {Adv. Opt. Photon.},
number = {2},
pages = {242--352},
publisher = {Optica Publishing Group},
title = {Integrated photonics on thin-film lithium niobate},
volume = {13},
month = {Jun},
year = {2021},
url = {https://opg.optica.org/aop/abstract.cfm?URI=aop-13-2-242},
doi = {10.1364/AOP.411024},
}

@article{Lu2026MultimodeSingleRing,
  title = {Multimode Single-Ring Photonic Molecule},
  author = {Lu, Jinsheng and Benea-Chelmus, Ileana-Cristina and Ginis, Vincent and Ossiander, Marcus and Shchepanovich, Danilo and Capasso, Federico},
  journal = {Phys. Rev. Lett.},
  volume = {136},
  issue = {10},
  pages = {103803},
  numpages = {7},
  year = {2026},
  month = {Mar},
  publisher = {American Physical Society},
  doi = {10.1103/vfbg-y973},
  url = {https://link.aps.org/doi/10.1103/vfbg-y973}
}

@article{Xu:21bragg,
author = {Yuntao Xu and Ayed Al Sayem and Chang-Ling Zou and Linran Fan and Risheng Cheng and Hong X. Tang},
journal = {Opt. Lett.},
number = {2},
pages = {432--435},
publisher = {Optica Publishing Group},
title = {Photorefraction-induced Bragg scattering in cryogenic lithium niobate ring resonators},
volume = {46},
month = {Jan},
year = {2021},
url = {https://opg.optica.org/ol/abstract.cfm?URI=ol-46-2-432},
doi = {10.1364/OL.414702},
}

@article{zhang2019broadband,
  title={Broadband electro-optic frequency comb generation in a lithium niobate microring resonator},
  author={Zhang, Mian and Buscaino, Brandon and Wang, Cheng and Shams-Ansari, Amirhassan and Reimer, Christian and Zhu, Rongrong and Kahn, Joseph M and Lon{\v{c}}ar, Marko},
  journal={Nature},
  volume={568},
  number={7752},
  pages={373--377},
  year={2019},
  publisher={Nature Publishing Group UK London}
}

@ARTICLE{1985ApPhA..37..191W,
       author = {{Weis}, R.~S. and {Gaylord}, T.~K.},
        title = "{Lithium niobate: Summary of physical properties and crystal structure}",
      journal = {Applied Physics A: Materials Science \& Processing},
         year = 1985,
        month = aug,
       volume = {37},
       number = {4},
        pages = {191-203},
          doi = {10.1007/BF00614817},
       adsurl = {https://ui.adsabs.harvard.edu/abs/1985ApPhA..37..191W}
}

@article{peterson1964electro,
  title={Electro-Optic Properties of LiNbO3},
  author={Peterson, GE and Ballman, AA and Lenzo, PV and Bridenbaugh, PM},
  journal={Applied Physics Letters},
  volume={5},
  number={3},
  pages={62--64},
  year={1964},
  publisher={American Institute of Physics}
}

@article{wang2018integrated,
  title={Integrated lithium niobate electro-optic modulators operating at CMOS-compatible voltages},
  author={Wang, Cheng and Zhang, Mian and Chen, Xi and Bertrand, Maxime and Shams-Ansari, Amirhassan and Chandrasekhar, Sethumadhavan and Winzer, Peter and Lon{\v{c}}ar, Marko},
  journal={Nature},
  volume={562},
  number={7725},
  pages={101--104},
  year={2018},
  publisher={Nature Publishing Group UK London}
}

@article{zhang2017monolithic,
  title={Monolithic ultra-high-Q lithium niobate microring resonator},
  author={Zhang, Mian and Wang, Cheng and Cheng, Rebecca and Shams-Ansari, Amirhassan and Lon{\v{c}}ar, Marko},
  journal={Optica},
  volume={4},
  number={12},
  pages={1536--1537},
  year={2017},
  publisher={OSA}
}

@article{zhu2022spectral,
  title={Spectral control of nonclassical light pulses using an integrated thin-film lithium niobate modulator},
  author={Zhu, Di and Chen, Changchen and Yu, Mengjie and Shao, Linbo and Hu, Yaowen and Xin, CJ and Yeh, Matthew and Ghosh, Soumya and He, Lingyan and Reimer, Christian and others},
  journal={Light: Science \& Applications},
  volume={11},
  number={1},
  pages={327},
  year={2022},
  publisher={Nature Publishing Group UK London}
}

@article{wang2023quantum,
  title={Quantum frequency conversion and single-photon detection with lithium niobate nanophotonic chips},
  author={Wang, Xina and Jiao, Xufeng and Wang, Bin and Liu, Yang and Xie, Xiu-Ping and Zheng, Ming-Yang and Zhang, Qiang and Pan, Jian-Wei},
  journal={npj Quantum Information},
  volume={9},
  number={1},
  pages={38},
  year={2023},
  publisher={Nature Publishing Group UK London}
}

@article{Jiang:17,
author = {Haowei Jiang and Rui Luo and Hanxiao Liang and Xianfeng Chen and Yuping Chen and Qiang Lin},
journal = {Opt. Lett.},
number = {17},
pages = {3267--3270},
publisher = {Optica Publishing Group},
title = {Fast response of photorefraction in lithium niobate microresonators},
volume = {42},
month = {Sep},
year = {2017},
url = {https://opg.optica.org/ol/abstract.cfm?URI=ol-42-17-3267},
doi = {10.1364/OL.42.003267},

}

@article{Hou_2024,
author = {Hou, Jiankun and Zhu, Jiefu and Ma, Ruixin and Xue, Boyi and Zhu, Yicheng and Lin, Jintian and Jiang, Xiaoshun and Chen, Xianfeng and Cheng, Ya and Ge, Li and Zheng, Yuanlin and Wan, Wenjie},
title = {Subwavelength Photorefractive Grating in a Thin-Film Lithium Niobate Microcavity},
journal = {Laser \& Photonics Reviews},
volume = {18},
number = {8},
pages = {2301351},
doi = {https://doi.org/10.1002/lpor.202301351},
url = {https://onlinelibrary.wiley.com/doi/abs/10.1002/lpor.202301351},
eprint = {https://onlinelibrary.wiley.com/doi/pdf/10.1002/lpor.202301351},
year = {2024}
}

@article{Lampert2025,
  author       = {Lampert, Yazan and Shams-Ansari, Amirhassan and Gaier, Aleksei and Tomasino, Alessandro and Cao, Xuhui and Magalhaes, Leticia and Rajabali, Shima and Lon{\v{c}}ar, Marko and Benea-Chelmus, Ileana-Cristina},
  title        = {Photonics-integrated terahertz transmission lines},
  journal      = {Nature Communications},
  year         = {2025},
  volume       = {16},
  number       = {1},
  pages        = {7004},
  doi          = {10.1038/s41467-025-62267-y},
  url          = {https://doi.org/10.1038/s41467-025-62267-y},
  issn         = {2041-1723}
}

@Article{nano8110910,
AUTHOR = {Wu, Rongbo and Wang, Min and Xu, Jian and Qi, Jia and Chu, Wei and Fang, Zhiwei and Zhang, Jianhao and Zhou, Junxia and Qiao, Lingling and Chai, Zhifang and Lin, Jintian and Cheng, Ya},
TITLE = {Long Low-Loss-Litium Niobate on Insulator Waveguides with Sub-Nanometer Surface Roughness},
JOURNAL = {Nanomaterials},
VOLUME = {8},
YEAR = {2018},
NUMBER = {11},
ARTICLE-NUMBER = {910},
pages = {910},
URL = {https://www.mdpi.com/2079-4991/8/11/910},
PubMedID = {30404137},
ISSN = {2079-4991},
DOI = {10.3390/nano8110910}
}

@article{Khalatpour2025,
author = {Khalatpour, Ali and Qi, Luke and Fejer, Martin M. and Safavi-Naeini, Amir H.},
title = {Roughness-Limited Performance in Ultra-Low-Loss Lithium Niobate Cavities},
journal = {Advanced Optical Materials},
volume = {14},
number = {8},
pages = {e02355},
doi = {https://doi.org/10.1002/adom.202502355},
url = {https://advanced.onlinelibrary.wiley.com/doi/abs/10.1002/adom.202502355},
eprint = {https://advanced.onlinelibrary.wiley.com/doi/pdf/10.1002/adom.202502355},
year = {2026}
}

@article{Zhu2024,
author = {Xinrui Zhu and Yaowen Hu and Shengyuan Lu and Hana K. Warner and Xudong Li and Yunxiang Song and Let\'{i}cia Magalh\~{a}es and Amirhassan Shams-Ansari and Andrea Cordaro and Neil Sinclair and Marko Lon\v{c}ar},
journal = {Photon. Res.},
number = {8},
pages = {A63--A68},
publisher = {Optica Publishing Group},
title = {Twenty-nine million intrinsic Q-factor monolithic microresonators on thin-film lithium niobate},
volume = {12},
month = {Aug},
year = {2024},
url = {https://opg.optica.org/prj/abstract.cfm?URI=prj-12-8-A63},
doi = {10.1364/PRJ.521172},
}

@Article{Arnold2025,
author={Arnold, Georg
and Werner, Thomas
and Sahu, Rishabh
and Kapoor, Lucky N.
and Qiu, Liu
and Fink, Johannes M.},
title={All-optical superconducting qubit readout},
journal={Nature Physics},
year={2025},
month={Mar},
day={01},
volume={21},
number={3},
pages={393-400},
issn={1745-2481},
doi={10.1038/s41567-024-02741-4},
url={https://doi.org/10.1038/s41567-024-02741-4}
}

@Article{Warner2025,
author={Warner, Hana K.
and Holzgrafe, Jeffrey
and Yankelevich, Beatriz
and Barton, David
and Poletto, Stefano
and Xin, C. J.
and Sinclair, Neil
and Zhu, Di
and Sete, Eyob
and Langley, Brandon
and Batson, Emma
and Colangelo, Marco
and Shams-Ansari, Amirhassan
and Joe, Graham
and Berggren, Karl K.
and Jiang, Liang
and Reagor, Matthew J.
and Lon{\v{c}}ar, Marko},
title={Coherent control of a superconducting qubit using light},
journal={Nature Physics},
year={2025},
month={May},
day={01},
volume={21},
number={5},
pages={831-838},
issn={1745-2481},
doi={10.1038/s41567-025-02812-0},
url={https://doi.org/10.1038/s41567-025-02812-0}
}

@article{
AliMiri2019,
author = {Mohammad-Ali Miri  and Andrea Alù },
title = {Exceptional points in optics and photonics},
journal = {Science},
volume = {363},
number = {6422},
pages = {eaar7709},
year = {2019},
doi = {10.1126/science.aar7709},
URL = {https://www.science.org/doi/abs/10.1126/science.aar7709},
eprint = {https://www.science.org/doi/pdf/10.1126/science.aar7709}}

@Article{Thompson2008,
author={Thompson, J. D.
and Zwickl, B. M.
and Jayich, A. M.
and Marquardt, Florian
and Girvin, S. M.
and Harris, J. G. E.},
title={Strong dispersive coupling of a high-finesse cavity to a micromechanical membrane},
journal={Nature},
year={2008},
month={Mar},
day={01},
volume={452},
number={7183},
pages={72-75},
issn={1476-4687},
doi={10.1038/nature06715},
url={https://doi.org/10.1038/nature06715}
}

\end{document}